\documentclass{aa}  

\usepackage[usenames,dvipsnames]{xcolor}
\usepackage{graphicx}
\usepackage{txfonts}
\usepackage{colortbl}
\usepackage{float}
\usepackage[colorlinks = true,
            linkcolor = blue,
            urlcolor  = blue,
            citecolor = blue,
            anchorcolor = blue]{hyperref}

\newcommand{\FeH}{[\mathrm{Fe/H}]}

\newcommand{\rrab} {\mbox{RR\emph{ab}}}

\newcommand{\typeab} {\mbox{\emph{ab}}}

\newcommand{\Gaia} {\textit{Gaia}}
\newcommand{\RRab} {\textit{RRab}}

\newcommand{\gapzo}{\emph{GAPZO}}

\def\newv2#1{{\textcolor{black}{#1}}}

\begin{document}

\title{First direct detection of an RR Lyrae star conclusively associated with an intermediate-age cluster}


   \author{Cecilia Mateu
          \inst{1} \thanks{cmateu@fcien.edu.uy}
          \and
          Bolivia Cuevas-Otahola\inst{2} 
          \and 
          Juan Jos\'e Downes\inst{1}
          }

   \institute{$^1$ Departamento de Astronom\'{i}a, Instituto de F\'{i}sica, Universidad de la Rep\'{u}blica, Igu\'a 4225, CP 11400 Montevideo, Uruguay\\
   $^2$ Instituto de Astronom\'ia, Universidad Nacional Autonoma de M\'exico, Ciudad Universitaria, Coyoac\'an, CP 04510, Ciudad de M\'exico, M\'exico\\
         }

   \date{Received September 15, 1996; accepted March 16, 1997}

  \abstract
{RR Lyrae stars have long been considered unequivocal tracers of old (>10 Gyr) and metal-poor ($\FeH<-0.5$) stellar populations. 
First, because these populations are where they are readily found and because, according to canonical stellar evolution models for isolated stars, these are the only populations where RR Lyrae should exist.
Recent independent results, however, are challenging this view and pointing at the existence of intermediate-age RR Lyrae, only a few (2--5) Gyrs old.}
{Our goal in this work is to provide direct evidence of the existence of intermediate-age RR Lyrae by searching for these stars in Milky Way open clusters, where the age association will be direct and robust. }
{We searched a catalogue of over 3,000 open clusters with published kinematically associated member stars by crossmatching it against a compilation of the largest publicly available RR Lyrae surveys (\Gaia, ASAS-SN, PanStarrs1, Zwicky Transient Facility and OGLE-IV).  }
 {We identified a star as a bona fide RR~Lyrae variable and robust member of the 2--4~Gyr old Trumpler 5 cluster, based on its parallax and proper motions and their agreement with confirmed cluster members. We derived an extremely low probability ($0.049\pm 0.013$\%) that the star is a background field RR~Lyrae and provide initial constraints on a possible binary companion based on its position in the colour-absolute magnitude diagram.}
{Currently a source of debate, the Trumpler~5 RR Lyrae provides the most direct evidence to date of the existence of RR Lyrae stars at much younger ages than traditionally expected and adds to the mounting evidence supporting their existence.}

   \keywords{ Stars: variables: RR Lyrae --
              Stars: evolution --
              Galaxy: stellar content -- 
              Galaxy: open clusters and associations
               }

   \maketitle
%

\section{Introduction}\label{s:intro}

RR Lyrae (RRL) stars are classical pulsator core He-burning stars produced at the intersection of the instability strip and the horizontal branch (HB). A convenient combination of observational properties have made them an extremely popular tracer of Milky Way (MW) structure in the current era of deep, large-scale, synoptic photometric surveys. Due to their distinctive variability they are easily and reliably identified based on (multi-epoch) photometry alone; and their standard candle nature provides individual distances with excellent precisions (down to a few percent), which, coupled with their large luminosities have made them ideal tracers of the tridimensional structure of our Galaxy and its surroundings \citep[see e.g.][]{Zhang2025,Prudil2025,CabreraGadea2025,Ngeow2025,Medina2024,Dorazi2024}.

Recent studies are defying the canonical notion that RRLs are tracers of exclusively old and metal-poor populations \citep[e.g.][]{CatelanSmith2015,Smith1995}. In the first large scale study of RRL kinematics in the MW using \Gaia\ DR2 \citep{GaiaCol_2018_DR2_survey}, \citet{Iorio2021} found a significant population of metal-rich RRLs ($\FeH\gtrsim -0.5$ up to nearly solar) with kinematics resembling thin disc populations 2--5~Gyrs-old, much colder and  faster-rotating than observed for disc populations at $>10$~Gyrs, the age expected for canonical RRL. Nearly simultaneously, in an independent study, \citet{Sarbadhicary_2021} concluded that almost half ($51\%$) the RRLs in the Large Magellanic Cloud (LMC) are associated to intermediate-age populations with ages from $1.2$ to 8 Gyrs. Their study was based on the Delay Time Distribution (DTD) of RRLs, a measure of the number of objects per unit (initial) mass formed after a burst of star formation, inferred from the spatial distribution of RRLs and a map of the stellar age distribution across the LMC. Both studies, based on completely independent data and methods, pointed at the existence of intermediate-age populations of RRLs only a few Gyrs old, much younger --and in the case of the MW, also more metal-rich-- than could be explained by canonical stellar evolution models. 
More recent studies of the kinematics of RRLs in the Galactic disc's warp \citep{CabreraGadea2025} and of the correlation of the kinematics of RRL and Mira stars of different ages determined via  the Mira's period-age relation \citep{Zhang2025} have further supported the kinematical intermediate age of thin disc RRLs found by \citet{Iorio2021}.

Although these studies strongly support the existence of intermediate-age RRL stars, their conclusions rely on indirect age estimates and/or statistical inferences about the expected fraction of such stars. The only RRL so far identified as a `young' object is the bulge field star OGLE-BLG-RRLYR-02792 found by \citet{Pietrzynski2012}, and is also one of only two RRL robustly confirmed to be part of a binary system \citep{Hajdu2021}. The star has a current mass of 0.26~$M_\odot$ and is found to be 5.4~Gyrs old based on binary evolution models \citep{Pietrzynski2012,Smolec2013}. \citet{Pietrzynski2012} show this mass is too low for the star to be undergoing core He-burning, calling it an RRL impostor and the prototype of a new `Binary Evolution Pulsator' class.
These precedents highlight the need for a direct and unambiguous identification of individual RRLs associated with an intermediate-age cluster, where the age association is unequivocal and the non-variable population of the cluster can serve as a benchmark for atmospheric abundances to be compared against. Such objects will provide key constraints to the evolutionary channels that are being proposed to explain the existence of these `young' RRL, e.g. by stellar population synthesis models including binary evolution with mass transfer \citep{Karczmarek2017,Bobrick2024} or by single stellar evolution with increased mass loss \citep{Bono1997a,Bono1997b}.

In \citet{CuevasOtahola2025} we conducted a search for RRLs associated to intermediate-age clusters in the Magellanic Clouds (MCs), in an attempt to find direct evidence confirming the existence of these stars at such early ages. The search yielded 23 RRL possible members in 10 clusters with ages from 2 to 8 Gyrs. Given how distant the clusters are, the current precision of \Gaia~DR3 proper motions is not yet enough to offer conclusive kinematic memberships and about half the RRL are expected to be field contaminants from the MC's population, awaiting spectroscopic follow-up for radial velocities to confirm their memberships to the clusters. 

In the MW the vast majority of clusters found at ages younger than 8~Gyrs are open clusters (OCs) whose masses are typically $<10^3M_\odot$ \citep[with a handful of Young Massive Clusters having larger masses,][]{PortegiesZwart2010}. The present-day formation rate (PTF) of \(0.83\,\mathrm{RRLs}/10^{5}\,M_\odot\) found by \citet{CuevasOtahola2025} for clusters in that age range implies only 1 in a few thousand clusters is statistically expected to host a single RRL and a search over a large sample of several thousand clusters is warranted. Such a search has become feasible thanks to \textit{Gaia}, which has enabled millions of stars to be catalogued as members of a vast number of OCs (over 5,000) with kinematic membership probabilities robustly
estimated based on parallax and proper motions 
\citep[e.g.][]{Hunt2023,CantatGaudin2018,CantatGaudin2020}. 
A quick estimate from the cluster masses reported by \citet{Hunt2024} and the \citet{CuevasOtahola2025} inferred rates yields a total mass of 
$\sim4\times10^5M_\odot$ for the 172 clusters in the age range 
from 1 to 8 Gyrs, which implies $\sim2$--3 RRLs could be lurking 
in the population of known Galactic OCs.

In this work we present the first RRL star conclusively associated to an intermediate age population: the 2--4~Gyr-old cluster Trumpler~5. This provides direct and model-independent evidence of the age of this young RRL star. 
In Sec.~\ref{s:data} we describe our search, conducted with a compilation of public RRL surveys crossmatched against the catalogue of stellar members associated to OCs from \citet{Hunt2023}. Our search 
resulted in an initial sample of 15 candidates, out of which we confirm one star as a bona fide RRL. In Sec.~\ref{s:results_the_star} we show the star to be robustly identified as an RRL, and as a cluster member by comparing it against cluster member's parallax, kinematics and position in the colour-magnitude diagram (CMD), and by showing that it has an extremely low probability ($0.049 \pm 0.013\%$) of being a background field RRL. 
In Sec.~\ref{s:discussion} we discuss constraints on a possible binary companion  based on its position in the CMD and the spectral energy distribution (SED), and Trumpler-5 as a cluster which favours stellar evolution with mass transfer.
We present our conclusions in Sec.~\ref{s:conclusion}.


\section{Data}\label{s:data}

In this work we use the OC census from \citet{Hunt2023} with mean cluster parameters from their Table~3 and members data from Appendix~A of \citet{Hunt2024}. In this latter work, the census from \citet{Hunt2023} was updated by including colour-magnitude diagram information in the cluster classification scheme and reporting extinction, age and distance based on isochrone fitting, as well as Jacobi radius and stellar mass for each cluster. 

For the RRLs we use the catalogue described in \citet{CabreraGadea2025} which combines the \Gaia~DR3 SOS catalogue from \citet{Clementini2023_SOS_GaiaDR3}, the PanSTARRS1 (PS1) catalogue from \citet{Sesar2017} and the ASAS-SN from \citet{Jayasinghe2019a,Jayasinghe2019b}, augmented with crossmatches against the Zwicky Transient Facility (ZTF) and the Optical Gravitational Lensing Experiment (OGLE) IV catalogues from \citet{Chen2020} and \citet{Soszynski2016}, respectively, following the crossmatching strategy described in \citet{CabreraGadea2025}. The final \gapzo\ catalogue contains a total of 309,998 RRL stars spanning the full sky.

For our analysis we crossmatched the \gapzo\ RRL catalogue by \Gaia~DR3 \verb|source_id| against the OC member table from \cite{Hunt2024}, keeping the 3,530 clusters classified strictly as OCs (Type=`o') to avoid globular clusters and moving groups, and with \verb|CMDCl50|$>=0.5$ and \verb|CST|$>5$ in order to keep only reliable OCs as suggested by \citet{Hunt2023}, and limiting our search to members within each cluster's tidal radius. This results in a sample of 3,358 young ($<1$~Gyr) and 172 intermediate age clusters (1--8~Gyr), in which an initial number of 15 RRLs were identified as members of the same number of OCs. Out of these, 13 stars were identified as RRL by a single survey, one by two surveys (ZTF and \Gaia) and the last one by all five surveys. Based on visual inspection of their light curves and position in the CMD, 14 out of the 15 stars were discarded based on their light curves (the majority were reclassified as eclipsing binaries) and/or their position in the CMD, as described in detail in Appendix~\ref{a:discarded}. In what follows, our discussion will focus on the last standing RRL star.

\section{Results: The RRL star in Trumpler~5}\label{s:results_the_star}

\begin{table}
\caption{Astrometric and physical parameters for the Trumpler~5 cluster and the RRL star}             
\label{t:astrom}      
\centering                          
\begin{tabular}{l l}      
\hline\hline                 
\multicolumn{2}{c}{Trumpler 5 Cluster} \\
\hline
(RA,DEC)          & (99.12705764, +9.45854713) deg\\
($l$,$b$)         & (202.82, +1.02) deg \\
Parallax          & (0.302$\pm$ 0.002) mas \\
pmRA$^*$          & (-0.618$\pm$ 0.003) mas/yr \\
pmDEC             & (0.269$\pm$ 0.002) mas/yr \\
Mass              & $(2.46 \pm +0.17)\times 10^4$ $M_\odot$ \\
Age              & $2.2-4.2$ Gyr \\
                 & 2.5~Gyr (O25) \\
                 & ($2.17_{-0.70}^{+1.00}$) Gyr (H23)\\
                 & ($3.5 \pm 1.7$)~Gyr (D21) \\
                 &  $4.2$~Gyr (CG20) \\ 
Distance          & ($2.905_{-0.009}^{+0.010}$) kpc (H23) \\
                  &  3.047~kpc (CG20) \\
$\FeH$            & $-0.403 \pm 0.006$ (D15) \\
 $A_V$             & ($1.71 \pm +0.19$) mag (H23) \\
$r_t$             & ($0.50$ deg, 25.5 pc) (H23)\\
$r_c$             & ($0.077$ deg, 3.9 pc) (H23)\\
$r_{50}$          & ($0.11$ deg, 5.4 pc) (H23)\\
\hline
\multicolumn{2}{c}{RRL} \\
\hline
\verb|source_id|  & 3326852328563919744 \\
(RA,DEC)     & (98.90796904,9.72592814) deg\\
($l$,$b$)    & (202\fdg48,0\fdg95) deg\\
Parallax     & (0.306 $\pm$ 0.028) mas \\
Distance (trig.) & $2.95 \pm 0.27$ kpc (from parallax) \\
$A_V$  & ($1.65 \pm 0.15$) mag (G19) \\    
         & ($1.503 \pm 0.090$) mag (L22) \\     
pmRA$^*$     & (-0.627 $\pm$ 0.031) mas/yr \\
pmDEC        & (0.328 $\pm$ 0.026) mas/yr \\
$G$          & (14.874 $\pm$ 0.012) mag \\
\verb|ruwe|       & 1.1 \\
BEP               & 1.3 \\
\hline                                   
\end{tabular}
\tablefoot{Reference legend: O25=\citet{Ozdemir2025}, H23=\citet{Hunt2023}, D21=\citet{Dias2021}, CG20=\citet{CantatGaudin2020}, D15=\citet{Donati2015}, G19=\citet{Green2019}, L22=\citet{Lallement2022}.}
\end{table}

Our search resulted in one star robustly identified as an RRL and as a member of an intermediate-age OC, the Trumpler~5 cluster. The RRL is identified as \verb|source_id|=3326852328563919744 by \Gaia\ DR3. 
\Gaia\ DR3 astrometric parameters \citep{GaiaCol_DR3} for the RRL and physical and astrometric parameters for the cluster from several sources are summarised in Table~\ref{t:astrom}. 

\citet{Anderson2025} recently conducted a similar search for variable stars in the \citet{Hunt2023} OC census by crossmatching against the \Gaia\ DR3 Specific Objects Study (SOS) catalogues for different classes of variable stars and found two stars classified as RRLs by \Gaia. Upon inspection of their light curves and position on the CMD, the stars turned out to be likely eclipsing contact binaries and, consequently, the authors concluded no RRLs were found associated to OCs.  Their search, however, was motivated by their study of much less luminous classes of variables and their sample limited to clusters nearer than 2.5~kpc, a limitation we did not need to impose here.

\subsection{A bona fide RRL}

   \begin{figure*}
   \centering
   \includegraphics[width=\textwidth]{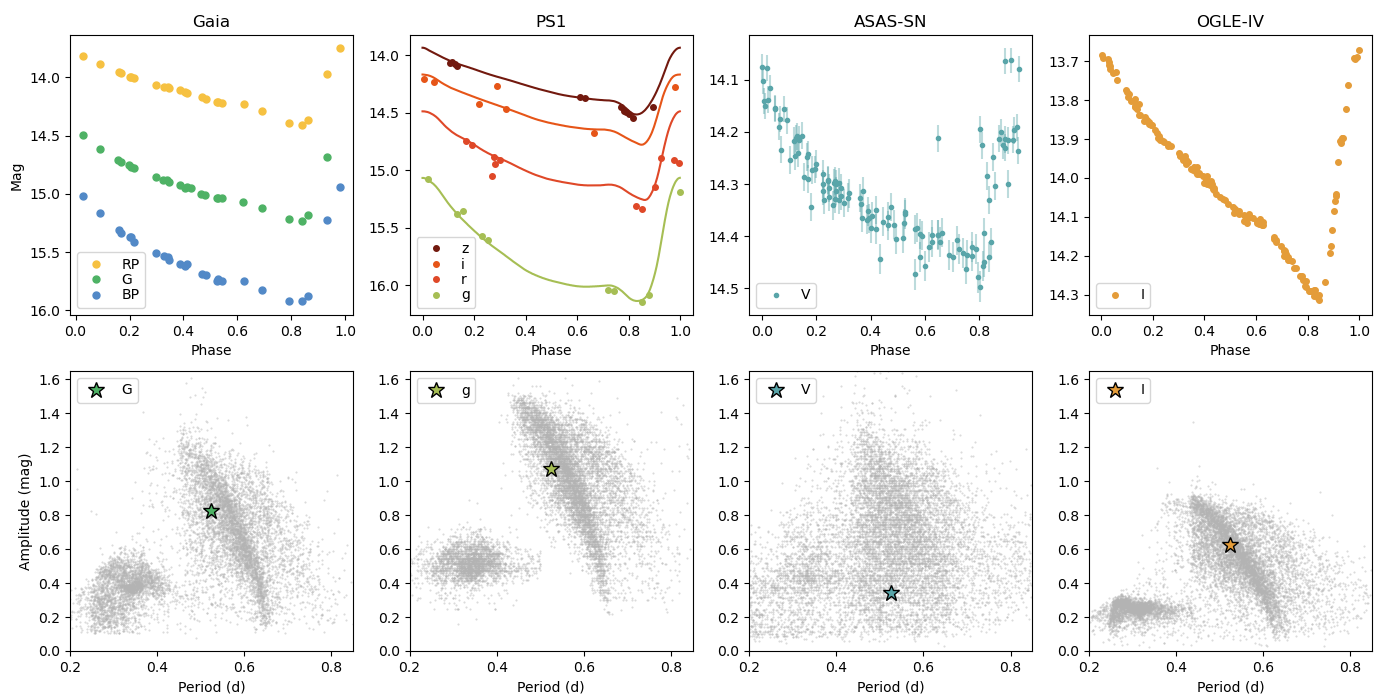}
   \caption{ Phase-folded light curves \emph{(top row)} and location of the Trumpler~5 RRL in the Period-Amplitude diagram \emph{(bottom row)} for the four surveys with publicly available time series data, from left to right: \Gaia\ \citep{Clementini2023_SOS_GaiaDR3}, PS1 \citep{Sesar2017c}, ASAS-SN \citep{Jayasinghe2019a} and OGLE-IV \citep{Soszynski2016}. In the bottom row the location of the Trumpler~5 RRL (star symbol) is shown in comparison to RRLs from each survey in the selected band. See text for details. Error bars for the magnitudes are plotted in all panels in the top row but are smaller than the symbol size in all cases except for ASAS-SN. }\label{f:lcs}%
    \end{figure*}

The star is consistently reported as an \rrab\ in all five surveys in the \gapzo\ catalogue. Table~\ref{t:P_Ampl} summarises its main light curve parameters and Figure~\ref{f:lcs} shows the star's light curves (top row) in the 9 different filters for the 4 surveys with publicly available time series data. In all cases the light curves show the typical sawtooth shape expected from its classification as an RRL of type \typeab. As observed in Figure~\ref{f:lcs} and Table~\ref{t:P_Ampl}, amplitudes decrease as filters become redder, as expected for a pulsating star \citep[e.g.][]{Kinman2010,CatelanSmith2015}, except for the V-band amplitude from ASAS-SN. The location of the star in the Period-Amplitude diagram (bottom row) for all bands is consistent with the expectation from the Period-Amplitude anti-correlation in all 8 bands available for \Gaia, PS1 and OGLE-IV (only one band per survey is shown). For ASAS-SN the relatively low V-band amplitude places the star below the main locus, however, the overall distribution for ASAS-SN RRL is much more scattered than usually observed, which is also reflected in a much weaker Period-Amplitude anti-correlation \citep{CatelanSmith2015}, in turn leading to lower amplitudes being more common in this catalogue. This suggests the low amplitude is more likely due to ASAS-SN amplitudes having a tendency to be underestimated, rather than it having a more physical explanation like the RRL being a Blahzko star, which may have explained its lower amplitude. We therefore conclude the star's observed light curve, period and amplitudes are typical for an \rrab.

\begin{table}
\caption{Period, light curve amplitudes and (intensity averaged) mean magnitudes for the Trumpler~5 RRL}             
\label{t:P_Ampl}      
\centering                          
\begin{tabular}{c c c c c}        
\hline\hline                 
Survey & Period & Band & Amplitude & MeanMag \\    
       &(d)  &      & (mag)     & (mag)\\
\hline                        
\Gaia\ SOS &  0.524806 & $BP  $ & 1.07 & 15.50 \\
         &           & $G   $ & 0.82 & 14.87 \\
         &           & $RP  $ & 0.69 & 14.09 \\
\hline
ASAS-SN  &  0.524830 & $V   $ & 0.34 & 14.31 \\
\hline
PS1      &  0.524815 & $g   $ & 1.07 & 15.84 \\
         &           & $r   $ & 0.77 & 15.00 \\
         &           & $i   $ & 0.61 & 14.53 \\
         &           & $z   $ & 0.58 & 14.28 \\
\hline
ZTF      &  0.524790 & $r   $ & 0.81 & 14.88 \\
\hline
OGLE     &  0.524807 & $I   $ & 0.62 & 13.99 \\

\hline                                   
\end{tabular}
\tablefoot{References are those provided for each survey in Sec. \ref{s:data}. }
\end{table}

\subsection{Kinematic Cluster Membership}
   \begin{figure*}
   \centering
   \includegraphics[width=\textwidth]{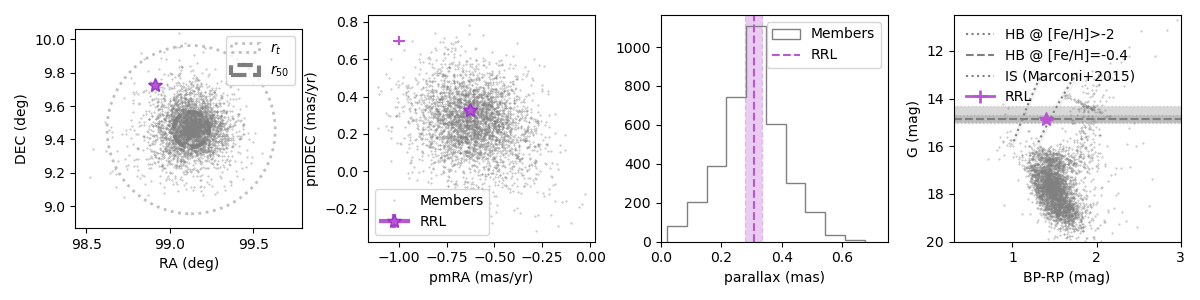}
   \caption{The Trumpler~5 RRL (star) and cluster members (light gray) are shown, from left to right, in the plane of the sky (DEC versus RA), proper motions (pmDEC versus pmRA), parallax and \Gaia\ CMD. In the first panel, $r_t$ and $r_{50}$ are shown with the dotted and dashed lines, these correspond respectively to the tidal radius and radius enclosing 50\% of the members according to \citet{Hunt2023}. In the fourth panel, the dotted lines correspond to instability strip (IS) limits from \citet{Marconi2015} for the cluster's metallicity; the dashed line shows the expected apparent G-band magnitude for the HB (from the empirical PLZ relation) at the cluster's metallicity \citep[$\FeH=-0.4$,][]{Donati2015}, and the star's parallax distance and extinction according to \citet{Green2019}; the shading represents the uncertainty in the HB luminosity corresponding to the full range of (observed) RRL metallicities $\FeH \in [0.,2.0]$~dex and the uncertainties in the apparent magnitude and colour of the RRL are smaller than the star symbol. The agreement with the RRL's apparent magnitude is remarkable and the star lies right at the edge of the limits of the IS (see discussion in Sec.~\ref{s:camd}). }\label{f:5d}%
    \end{figure*}

Figure~\ref{f:5d} shows the RRL star's position relative to cluster members from \citet{Hunt2024}, from left to right, in the sky and proper motion planes, in a parallax histogram and in the colour-magnitude diagram (CMD), using \Gaia\ bands. As shown by the figure, the RRL lies within the cluster's tidal radius (by construction of our search) and its proper motions and parallax are in excellent agreement with the cluster's, supporting its likely membership at face value. 

Our Trumpler~5 RRL was identified by \citet{Hunt2023} as a member of the cluster with a membership probability $p=0.51$. The star had also been previously identified as a member of Trumpler~5 by \citet{CantatGaudin2018, CantatGaudin2020} using Gaia~DR2  and \citet{vanGroeningen2023} using Gaia~DR3, with probability 1 in all cases. All of these works compute membership probabilities by modelling the cluster in the same space of observables (sky, parallax and proper motions) with a key difference being that membership probabilities from  \citet{Hunt2023} were computed using HDBSCAN. According to the authors their algorithm does not take into account observational errors and bases the assignment of probabilities on the proximity of a given star to the bulk of the cluster's population, in contrast to the works of \citet{CantatGaudin2018,CantatGaudin2020} and \citet{vanGroeningen2023} in which observational uncertainties are accounted for, which may explain the much lower membership probability found for this star by \citet{Hunt2023}. Since this point is key, in Sec.~\ref{s:not_field} we revisit it and show the probability that the star may be a chance interloper of the thin disc field population is extremely low. 

Finally, an exhaustive search of major spectroscopic survey databases (DESI, LAMOST, SDSS-IV) as well as using Simbad and VizieR yielded no available measurements for the star's line-of-sight velocity in the literature.

\subsection{Metallicity and location in the CMD}

The star's photometric metallicity, estimated from the Fourier light curve $\phi_{31}$ parameter, has been reported by \citet{Clementini2023_SOS_GaiaDR3} as $\FeH=-0.44 \pm 0.34$, by \citet{Li2023} as $-0.86 \pm 0.33$, by \citet{Muraveva2025} based on the Gaia DR3 G-band, and as $-0.98\pm0.45$ and $-0.71\pm0.22$ by \citet{He2025} from ZTF $g$ and $i$ bands.
The dispersion between the measurements, particularly of those inferred from the same data, shows that photometric metallicity estimates are still quite imprecise in an individual basis. In the last three cases, although slightly more metal-poor than the cluster's, all estimates are deviated by $\leq1.4\sigma$ from the cluster's spectroscopic metallicity of $\FeH=-0.4$ \citep{Donati2015}. We therefore take the RRL's metallicity as being consistent with the cluster's within its uncertainties. 

The RRL's position in the CMD is also shown in Figure~\ref{f:5d} (last panel), where good agreement is found between the star's apparent $G$-band magnitude and that expected for the HB assuming the cluster's metallicity of $\FeH=-0.403\pm0.006$ \citep{Donati2015}, its distance and extinction $A_V=1.65$ from \citet{Green2019} and assuming the (empirical) absolute magnitude versus metallicity relation for the $G$ band from \citet[][Eq. 19]{Garofalo2022}. The IS limits for the first overtone blue edge (FOBE) and fundamental red edge (FRE) according to \citet{Marconi2015} for the cluster's metallicity are also shown. The limits provided by the authors in their Eq. 3 and 4 for $\log{T_\mathrm{eff},\log{L/L_\odot}}$ were transformed to the Gaia photometric system using the colour-$T_\mathrm{eff}$ transformation for Giants provided by \citet{Mucciarelli2021} in their Table~1 and a bolometric correction $BC_G$ for the G-band at $T_\mathrm{eff}=6500$~K from the 2019 update of Table~5 from \citet{PecautMamajek2013}\footnote{Available at \href{http://www.pas.rochester.edu/~emamajek/EEM_dwarf_UBVIJHK_colors_Teff.txt}{this URL}}. 

Finally, we searched \Gaia~DR3 flags \verb|ipd_frac_multi_peak| and \verb|ipd_gof_harmonic_amplitude| for signs of potentially unresolved binarity \citep{Lindegren2018} but found none. Also the star's \verb|ruwe| and \verb|phot_bp_rp_excess_factor| (BEP) are normal as shown in Table~\ref{t:astrom}.  

\subsection{Not a background RRL star}\label{s:not_field}

Our argument that this a bona fide intermediate-age RRL star hinges on a reliable association to the cluster. A definitive association would require a measurement of the star's radial velocity, currently unavailable. In the mean time, we can assess how likely it is that our star is a chance interloper from the field population. 

\begin{figure}
\centering
\includegraphics[width=\columnwidth]{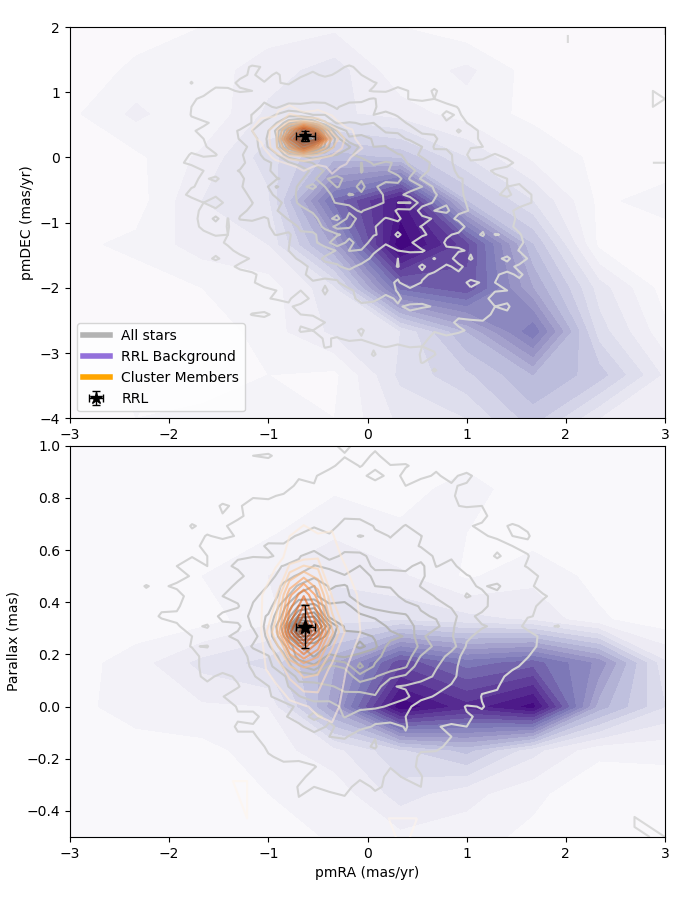}
\caption{Location of the Trumpler~5 RRL (black star) with respect to the field population in the proper motions (top panel) and parallax versus pmRA (bottom panel) planes. Iso-contours represent the density of cluster members (orange), all field stars (gray) within the cluster's tidal radius, and 2000 field RRL stars (violet) within in a FOV of 25\degr around the cluster.  A similar result is found when using proper motion in declination (not-shown). The analysis yields a final probability $(0.049 \pm 0.013)\%$ that the RRL is a chance interloper.
}
\label{f:bakground_model}
\end{figure}

We quantify the probability that a field MW RRL has proper motions and parallax 
compatible with our star's. For reference, within the cluster's projected tidal radius of $\sim0\fdg5$ there are only three other RRLs. Hence, a large 
window is necessary to include enough RRLs for the background to be modelled properly, 
which already hints the likelihood of chance interlopers is very low. We select a window 
such that at least 2000 RRL are included, which resulted in a radius of $\sim25^\circ$. 
Figure~\ref{f:bakground_model} shows the distribution of all field stars, field RRLs, cluster members and the RRL studied in this work. The distribution of field RRLs, although it includes  thin and thick disc contributions \citep[e.g.][]{Mateu2018,Prudil2020}, is 
dominated by halo RRLs, whose bulk kinematics differ significantly from the 
cluster's, which is kinematically associated to the thin disc. 

We use a Gaussian Mixture to model the 3D distribution of proper motions and parallax 
of the background RRLs. We draw 500K random  samples from the model and then compute the fraction whose proper motions and parallax coincide with the RRL star's to within 3 times its uncertainties and estimate the probability and its error as the mean and standard deviations from 100 bootstrap realisations. This yields a probability $p = (0.012 \pm 0.003)\%$. The final probability that such a star has been drawn at random in a sample of only four stars (the number of RRLs within the 
cluster's projected tidal radius) is given by the binomial 
probability of having one success in a sample of 4 draws when the probability 
of success is the $p$ found above. This yields a final probability of
$(0.049 \pm 0.013)\%$ that the RRL is a chance interloper. It 
is clear, then, the probability of the RRL in Trumpler~5 being a chance interloper is 
extremely low (1 in 2,000) and we take the association based on proper motions and 
parallax as a very robust one, even without the radial velocity information. 

\section{Discussion}\label{s:discussion}

\subsection{Formation scenarios} \label{s:formation_scenarios}

\begin{figure}
\centering
\includegraphics[width=\columnwidth]{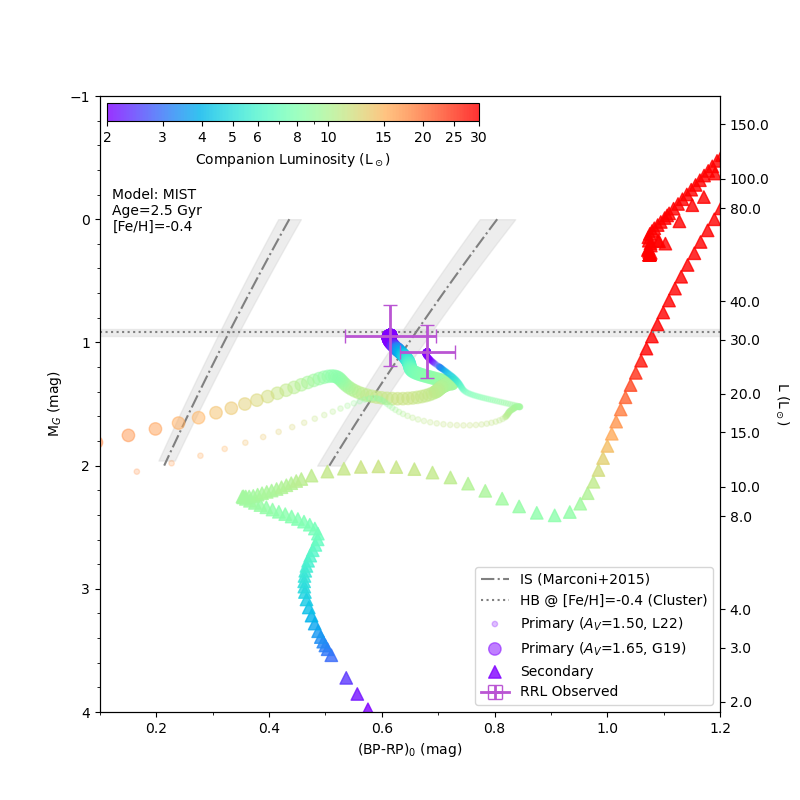}
\caption{CAMD for the cluster members (gray) and RRL (purple dots with error bars).  Possible primary RRL stars are shown as 
coloured circles, and the corresponding secondary companions as triangles. The two tracks shown for primaries correspond to the G19 (large circles) and L22 extinctions (small circles). The colour scale represents the mass of the companion star. The IS limits for \RRab~from \citet{Marconi2015} and expected HB luminosity for the cluster's metallicity from the \citet{Garofalo2022} PLZ relation (as in Fig.~\ref{f:5d}) are shown by the dotted lines, the shaded areas correspond to their respective uncertainties.}
\label{f:cmd_binarity}
\end{figure}

We have established that the RRL star is a robust member of the Trumpler~5 cluster. This provides a direct measurement of the RRL's age by virtue of its robust association to the cluster, independent on any assumptions on the nature and evolutionary formation channel.

The existence of an RRL star at such a young age (2--4~Gyr) challenges canonical stellar evolution channels and requires its progenitor to have lost a significant amount of  mass in order for the star to have entered the IS and become an RRL pulsator. 
Two main avenues have been proposed for this: increased mass loss in single-star evolution models \citep{Bono1997a,Bono1997b,Taam1976}; and binary interactions with mass transfer \citep{Bobrick2024,Iorio2021,Sarbadhicary_2021}.

\citet{Bono1997a,Bono1997b} explored theoretical single-stellar evolution scenarios to produce metal-rich RRL via increased mass loss. Accounting for the increase in He abundance with metal content and using different prescriptions of mass loss at the RGB, their models are able to reproduce the formation of metal-rich RRL up to solar metallicities for ages $>2.5$~Gyr and even younger for increased mass loss values. The zero-age HB luminosity found for these models  ($\log{L/L_\odot}\sim1.5$, $L\sim31L_\odot$) for ages $>2.5$~Gyr and half the solar metallicity, is in agreement with more recent PLZ values for such metal-rich RRL \citep[e.g.][see Fig. 4 below]{Garofalo2022}. The `cause' of the increased mass loss, however, is not exclusively determined by the He enhancement and different mass loss prescriptions are explored by the authors. For the Trumpler~5 RRL about 1~$M_\odot$ would need to have been lost between the RGB and the HB stage for the star to reach the core He-burning stage with a core mass $\sim0.5-0.6M_\odot$ reaching the IS. As discussed by \citet{Zhang2025} in their Sec.~5.2 and shown in their Fig.~9, such a high mass lost far exceeds the expectations from available wind mass loss estimates, which predict at most $\Delta M\sim0.3M_\odot$, based on metal-poor populations \citep{Gratton2010,Savino2020,Tailo2021,Howell2024} or even less for more metal-rich populations \citep{Miglio2021}. Additionally, the single stellar evolution effect responsible for such a high mass loss rate would need to have affected only the RRL, given that the red clump population of the cluster is found to be well described by stellar evolution models with normal prescriptions and He abundance \citep{Ozdemir2025}. Since these tensions disfavour the single stellar evolution scenario, we now turn our focus to the binary evolution scenario.

The second scenario is that of binary evolution, in which the Trumpler~5 RRL could either be a `normal' RRL, currently core He-burning after evolution undergoing mass-transfer, or it could be a Binary Evolution Pulsator or `RRL impostor' such as the \citet{Pietrzynski2012} object. The pulsation properties predicted by \citet{Smolec2013} for these objects (red open squares in their Fig. 1) are roughly consistent with our star's period and amplitude. The BEP's light curve, however, is more irregular and less clearly saw-tooth shaped than our star, which has the light curve of a canonical RRab (see Fig.~\ref{f:lcs}). The light curve dissimilarities do not seem to support this scenario, however, in order for a conclusive case to be made, models should be tailored to the specific properties of the Trumpler~5 RRL and, ideally, a dynamical estimate of the mass would be needed, which would require for the star to be in a double-lined spectroscopic binary.

The lack of spectroscopic observations means that there is currently no direct evidence confirming (or ruling out) binarity. We analysed the \textit{Gaia} \verb|ruwe| parameter, which is consistent with a single star; however, this does not rule out low-mass and/or wide companions to 
which \verb|ruwe| is not sensitive. The absence of visual companions only allows us 
to discard very wide binaries (an angular separation of $\sim1^{\prime\prime}$ 
corresponds to $\sim1000$~AU), which in any case would not be expected to 
had affected the evolution of the RRL.
We now attempt to constrain the properties of a possible unresolved companion using photometry 
in two ways. First, we analyse the colour--absolute-magnitude diagram (CAMD) to estimate what type of companion could 
explain the position of the RRL star, slightly outside the limits of the IS. Second, we explore the presence of excess emission in the spectral 
energy distribution (SED) that could arise from a companion or from circumstellar material.

\subsection{The binary evolution scenario}

\subsubsection{Location in the CAMD}\label{s:camd}

Figure~\ref{f:cmd_binarity} shows the location of the Trumpler~5 RRL star 
in the CAMD using \textit{Gaia} photometry. 
The absolute magnitude was computed from its intensity-averaged $G$-band 
magnitude from \textit{Gaia}~DR3 SOS \citep{Clementini2023_SOS_GaiaDR3}, 
its parallax distance of $2.95\pm0.27$~kpc computed including the $-0.033\pm0.002$~mas offset correction found by \citet{Garofalo2022} for RRL stars.
We have kept this analysis based on the stars' trigonometric distance inferred from its parallax, rather than a photometric distance based on an empirical Period-Luminosity-Metallicity (PLZ) relation, since this is a geometric distance completely independent of any assumptions on the star's nature as an RRL or of the validity of the PLZ for these unusual stars. 
Although metal-rich field RRL,  probably also young (as mentioned in Sec.~\ref{s:intro}) and likely similar to the Trumpler~5 RRL, have been used as part of many calibration samples of PLZ relations \citep[see e.g.][]{Muraveva2018,Garofalo2022,Prudil2025}, the extent of the validity of the PLZ relation for these objects is an aspect that remains to be explicitly tested. 
Hence, we opt to base our estimates of the RRL's location in the CAMD on its trigonometric (parallax-based) distance exclusively, allowing us to compare its derived luminosity (or absolute magnitude) to predictions from PLZs.

The extinction correction turns out to be the largest source of uncertainty in the calculation of both $M_G$ and colour because the cluster suffers from strong differential reddening \citep{Ozdemir2025}, which varies with distance and along each line of sight. The figure shows the results obtained assuming an extinction $A_V=1.65 \pm 0.15$ as predicted by the \citet{Green2019} map for the RRL's line of sight and its trigonometric distance, compared to an extinction of $A_V=1.50\pm0.09$ mag predicted by the \citet{Lallement2022} dust map (hereafter L22)\footnote{See also Fig.~\ref{fa:Av_vs_d} in Appendix~\ref{a:Av} for a more detailed comparison of L22 and G19.}. Extinction law coefficients from \citet{Fitzpatrick2019} were used for the conversion to the \textit{Gaia} bands. 
According to G19's extinction the star is inside the IS and its luminosity is in remarkable agreement with the HB luminosity expected from the \citet[hereafter G22][]{Garofalo2022} empirical Period-Luminosity-Metallicity (PLZ). According to L22's extinction, however, the star would be just past the red edge of the IS and slightly under-luminous compared to the HB from the G22 PLZ.

\begin{figure}
\centering
\includegraphics[width=\columnwidth]{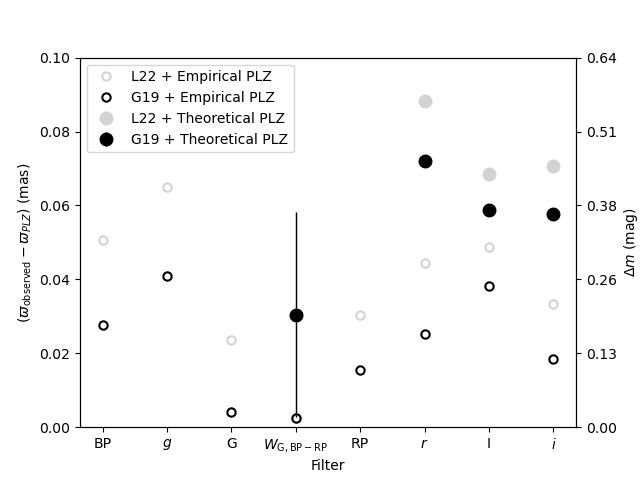}
\caption{Difference between the observed parallax and the parallax predicted from theoretical (filled circles) and empirical (open circles) PLZ relations for different photometric bands. The equivalent difference in magnitude is also represented in the right-hand axis.
Results for the predicted parallax are shown for assumed extinctions from the G19 (black) and L22 (gray) dust maps.
Since errors in this parallax difference are dominated by the uncertainty in the observed parallax, they are the same for all points and shown only for one band for guidance. 
Empirical PLZ relations used are from \citet[][Eq. 8]{Narloch2024} for PS1 bands, from \citet{Prudil2024} for $I$, $BP$ and $RP$ and from \citet{Garofalo2022} for the $G$ and $W_{G,BP-RP}$ band. Theoretical relations used are from \citet{Marconi2021} for $W_{G,BP-RP}$, \citet{Marconi2015} for $I$ and \citet{Marconi2022} for the LSST $r$ and $i$ bands (assumed here to match the PS1 bands, based on filter transmission curves for both surveys).  
}
\label{f:dm_vs_filter}
\end{figure}

Constraints on the extinction can be derived from a similar comparison made taking advantage of the photometric information available for the star in different filters (summarized in Table~\ref{t:P_Ampl}). Figure~\ref{f:dm_vs_filter} shows the difference between the observed parallax and the corresponding prediction from empirical (open) and theoretical (filled) PLZ relations, both for the G19 (black) and L22 (gray) extinctions. This is also expressed equivalently in terms of a difference in magnitude, as shown by the right-hand axis. The figure shows the prediction from the empirical G22 PLZ for the reddening-free Wesenheit magnitude $W_{G,BP-RP}$ is in remarkable agreement with the observed parallax. This result is also in best agreement with that obtained for the G band using the (larger) G19 extinction, instead of the L22 extinction which yields a systematic offset of $\gtrsim0.02$~mas. Therefore, in what follows, we adopt the G19 extinction for the star.

Figure~\ref{f:dm_vs_filter} also shows differences in the range 0.02--0.04~mas in the predicted parallax for empirical PLZs and systematically larger in the predictions from theoretical PLZs for all available photometric bands. 
If the Trumpler~5 RRL were not a real member of the cluster, but rather a normal field star either captured or simply coinciding in parallax and proper-motion with the cluster, the theoretically predicted PLZ parallax should be in agreement with the observed one. 
Of particular interest is the discrepancy found for the theoretical predictions for the $W_{G,BP-RP}$ index, first because it is reddening-free, but also because this PLZ was empirically validated by \citet{Marconi2021} using Gaia DR2 astrometry for field RRL stars with robust metallicity information. 
In this band, the predicted parallax is $\sim0.03$~mas smaller than the observed parallax\footnote{Observed parallaxes already include the zero-point correction from G22. 
Not including this correction would even increase this discrepancy.}. The discrepancy observed for the $r$, $I$ and $i$ bands is even larger $\gtrsim0.06$~mas (or $\sim20\%$ in parallax or distance). 
Although there may be caveats due to this analysis being limited to a single object, and there is still consistent within the uncertainties, the systematic disagreement of the theoretical PLZ predictions found over the different bands and the smaller disagreement shown by empirical PLZs --based on RRL samples including field stars likely of a similar nature to the Trumpler~5 RRL-- are further arguments pointing against the possibility of the star being a normal RRL posing as a cluster member.

\subsubsection{Constraints on binarity}

To derive empirical constraints on the possible binarity of the star, we use its position on the CAMD to explore what should hypothetical secondary companions be such that the combined flux of the system matches the observed star while keeping the primary inside the IS, as required for it to develop the characteristic RRL pulsation \citep[e.g.][]{Smith1995}. 
To do this, we draw a secondary component from the normal 
cluster population and compute the primary's fluxes needed to reproduce the position of the observed star in the CAMD. This exercise requires adopting an age for the cluster, which, as discussed by 
\citet{Ozdemir2025}, is uncertain when relying on photometry alone due to strong 
differential extinction and the resulting age--metallicity--extinction degeneracy. 
Using NIR spectroscopy, \citeauthor{Ozdemir2025} derived atmospheric parameters for several giant stars in the cluster, performed differential-reddening corrections, and obtained an age of $2.5$~Gyr using MIST isochrones \citep{Dotter2016,Choi2016}. They also presented a strong argument for this age: the clearly observed gap caused by the blueward hook in the isochrones near the main-sequence turn-off, a feature found only in populations this young. Therefore, in what follows we
adopt their estimate of $2.5$~Gyr for the cluster age. 

Figure~\ref{f:cmd_binarity} shows the resulting sequence of possible RRL primaries 
and their corresponding companions (triangles) that reproduce the observed position of the unresolved binary, according to the MIST models \citep{Dotter2016,Choi2016}, for the G19 extinction (large circles). The result for the L22 extinction is also shown for comparison (small circles). 

As the figure shows, main-sequence (MS) companions with luminosities $\lesssim5 L_\odot$ are faint enough for the flux to be entirely dominated by the RRL primary and leave the star either in its observed position or slightly redder, still within the IS. More luminous MS companions, subgiant branch (SGB) and lower luminosity red giant branch (RGB) companions up to $\sim12 L_\odot$ are ruled out as, being bluer, they would require a cooler/redder primary outside the IS. As luminosity increases on the RGB, companions in the range $\sim12-18L_\odot$ start to be red enough to require a bluer primary inside the IS. Any companion more luminous than $\sim18L_\odot$ can also be ruled out as it would require a primary bluer than the blue edge of the IS. Further restrictions could be imposed on the companions knowing the observed star is an \rrab~ and, thus, the primary must be on the red side of the IS. In addition, if the primary were to follow normal PLZ relations, the lower its luminosity the shorter its pulsation period would be, again making it inconsistent with the observed properties of the star. These conclusions still hold for the lower extinction predicted by the L22 map, with minor changes to the luminosities of the possible companions. This analysis, therefore, supports either a low luminosity ($\lesssim5 L_\odot$) MS companion or no companion at all are consistent with the observed colour and magnitude of Trumpler~5's RRL.

The MS-companion scenario would be consistent with predictions from \citet{Bobrick2024} who used MESA stellar-population synthesis models including binary interactions to reproduce the production of young and metal-rich RRL stars. In their simulations, binary-made RRL stars are formed from evolved stripped-RGB stars and have surviving MS companions  with masses between $0.65$ and $1.9\,M_\odot$ and luminosities $15\lesssim L/L_\odot\lesssim 35$. Initial masses for binary-made RRL stars are found to be in the range $0.95$--$2\,M_\odot$  with mild mass ratios between $1.0$ and $1.55$. Because the primary is the star  that becomes an RRL after ascending the RGB, its initial mass correlates with  the age of the population. Since the mass ratios do not span a very large range, \citet{Bobrick2024} point out that the secondary mass also roughly  correlates with age, with systems younger than 3~Gyr having companion masses larger than $1\,M_\odot$ and luminosities $L>1.5L_\odot$. This would be consistent with the Trumpler~5 RRL and G19 extinction and with luminosities at the higher end of the range predicted by \citet{Bobrick2024}. It should be noted, however, that alternative scenarios of binary-made RRL formed after the binary system merges during the common envelope phase are also possible (Mateu et al. in prep.). Quantitative predictions and the relative efficiency of these channels are being tested as they depend on binary evolution parameters difficult to be constrained such as the critical values for mass transfer stability, mass loss parameters, among others.

\begin{figure}
\centering
\includegraphics[width=\columnwidth]{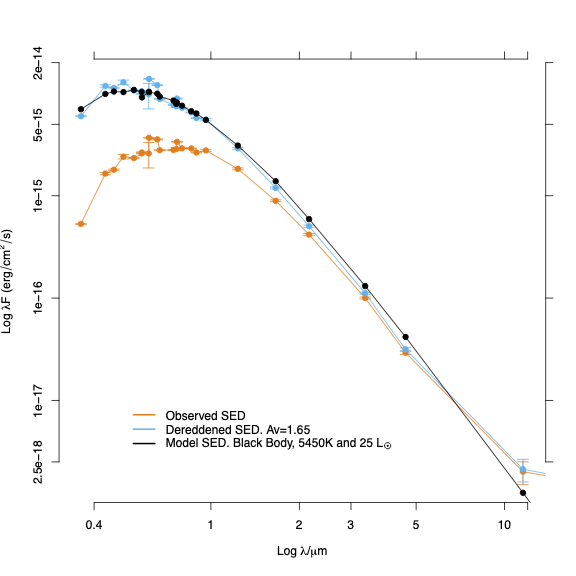}
\caption{Spectral energy distribution of the RRL best 
fitted to a black body with
$T_\mathrm{eff}=5450K$ and $L=25L_\odot$. The SED includes
magnitudes in 24 pass-bands from 2MASS, \Gaia~DR3, HST/ACS, 
IPHAS, PS1, SDSS and WISE surveys. The SED was dereddened
by $A_V=1.65$ and the extinction
law from \citet{Fitzpatrick1999}
and \citet{Indebetouw2005}.
}
\label{f:sed_binarity}
\end{figure}

Finally, we also explored the possible existence of an IR excess, based on the spectral energy distribution (SED) of the star. Figure \ref{f:sed_binarity} shows the dereddened SED  for the RRL star built using the VOSA tool \citep{Bayo2008} and available photometry from the literature over the wavelength range 
$0.36\lesssim\lambda/\mu m\lesssim 11$.
The  SED fits well to a single blackbody of $T_\mathrm{eff}=5450$~K with
a luminosity $L=25\,L_\odot$  is obtained when fitting only the optical wavelengths 
($\lambda\lesssim1.2\,\mu$m), and the same result is found when fitting over the full wavelength range. This result rules out a luminous RGB 
companion and also suggests the absence of IR excesses up to 
$\lambda\lesssim11\,\mu$m from circumstellar material.
The best fit to a two-blackbody model results in two indistinguishable 
components with the same temperature and half the luminosity each. Although the result supports the $T_\mathrm{eff}$ previously found, it is not sensitive to companions with luminosities well below that 
of the RRL star, as the resulting changes are comparable to the 
photometric uncertainties. We note that available interferometric observations with angular 
resolutions and sensitivity to resolve even faint companions at 
angular distances of tenths of $mas$ could probe for companion 
at distances from $30$ to $300\,AU$.

In summary, the observed RRL is with the luminosity of a normal RRL, based on expectations for the HB luminosity from existing empirical PLZ relations and inconsistent with predictions from theoretical PLZs. Its position in the CAMD would be consistent either with no companion or with MS companions with luminosities up to $\sim5L_\odot$, in agreement with predictions from the binary interaction models by \citet{Bobrick2024}.

\subsection{A binary-friendly cluster}\label{s:yss}

The Trumpler~5 RRL was also identified by \citet{Rain2021} as a member of this cluster using the OC members from \citet{CantatGaudin2018,CantatGaudin2020}, and classified it as a \emph{yellow straggler star} (YSSs), one of the two found in the cluster. In their work, YSSs are defined as stars redder than the main sequence turn-off, bluer than the red giant branch and more luminous than the subgiant branch, conditions indeed met by our star. In addition to the possibility of them being unresolved binaries whose combined fluxes place them in the YSS locus \citep{Rain2021,Mermilliod2007}, YSSs are of special interest since it has also been proposed they may have been formed by binary mergers, collisions and/or mass transfer from a red giant companion \citep{Landsman1997,Landsman1998,Leiner2016}.

\citet{Rain2021} also found 103 blue straggler stars (BSSs) in Trumpler~5, making it the OC with the largest absolute population of BSSs and the 8th in terms of its relative fraction, calculated as the fraction of BSSs relative to main sequence stars up to 1~mag below the turn-off. The correlation between the fraction of BSSs and the binary fraction of a population observed in open and globular clusters as well as in dwarf galaxies \citep[e.g.][]{Ferraro2026,Momany2015} implies Trumpler~5 hosts a relatively large population of binaries in comparison to other OCs.  

If binary evolution with mass transfer is indeed a viable evolutionary path for RRLs to be produced at intermediate ages, the relatively large mass of Trumpler~5 ($>24,000M_\odot$) combined with a comparatively large binary fraction could arguably `conspire' favourably for the cluster to have produced a binary-evolution RRL despite how inefficient this mechanism may be. 

It should also be noticed that Trumpler 5 is an atypical OC, being older and more metal-poor than typical OCs \citep{Ozdemir2025}, and with a mass larger than the canonical mass of OCs  \citep{PortegiesZwart2010}. Trumpler 5 is a dynamically evolved cluster as shown by an analysis of the profile of its BSSs \citep{Rain2021_dyn}, with a mass resembling that of a low-mass Young Massive Cluster (YMC), leading to think this could be a transition cluster between OC and YMC, considering OCs as the low-mass end of YMCs \citep{Bastian2016}.  

\section{Conclusions}\label{s:conclusion}

In this work we have provided the first conclusive association of an RRL to an intermediate-age star cluster: the 2--4 Gyr-old Trumpler 5. This provides the first direct measurement of an intermediate-age RRL's age, independent on any assumptions on the nature of the star or its possible evolutionary formation channel.

We have shown the star to be a bona fide RRL, identified independently and consistently by five RRL surveys: Gaia DR3 SOS, ASAS-SN, PS1, ZTF and OGLE-IV. Its full information, as well as the cluster's, is provided in Table~\ref{t:astrom}. The star was robustly identified as a member of the Trumpler 5 cluster by \citet{Hunt2024,CantatGaudin2018} and \citet{CantatGaudin2020}, it is located within the cluster's tidal radius, and its proper motions and parallax are in remarkable agreement with the cluster's members. Its observed apparent G-band magnitude is  consistent with the expectation from empirical RRL PLZ relations for the HB at the cluster's metallicity and at the star's distance and extinction according to the \citet{Green2019} dust map. Based on the parallax and proper motion distribution of field RRL stars and the total number of field RRLs observed within the cluster's tidal radius (3), we have shown the RRL to have a probability of $0.049\pm0.013\%$ of randomly sharing the cluster's parallax and proper motion to within 3 times the star's uncertainties, implying a very low chance of the RRL being a chance interloper. Given the size of the cluster sample (>3K OCs) searched \citep{Hunt2024}, having observed one RRL in the 1--8~Gyr age range implies a frequency of $0.25\times 10^{-5} M_\odot$~RRL which  agrees, to within Poisson noise, with the $\sim2$~RRL expected from the $0.2\pm0.2 \times 10^{-5} M_\odot$ PTF results obtained by \citet{CuevasOtahola2025} from LMC and SMC clusters, and with the $\sim0.1 \times 10^{-5} M_\odot$ rate predicted by binary evolution models from \citet{Bobrick2024}.  

Finally, although not firmly conclusive due to the large extinction uncertainties, we explored possible constraints on a binary companion based on the CAMD and the star's SED. 
With the assumed extinction ($A_V=1.65$) from the G19 dust map, the star lies inside the IS and has a luminosity consistent with predictions from empirical PLZ relations for the G and $W_{G,BP-RP}$ bands. Theoretical PLZs are found to predict systematically smaller parallaxes than observed, and also compared to predictions from empirical PLZs in all available bands, also disfavouring the possibility of the star being a normal background star posing as a cluster member.
Given its position in the CAMD, the RRL is consistent with being a single star or, if in a binary, the most likely scenario would require for the secondary to be a low luminosity MS star ($L<5L_\odot$). SGB or low luminosity RGB companions are also possible, but would require a significantly under-luminous RRL. Finally, the star's SED analysed with VOSA showed no signs of an IR excess from circumstellar material up to $11\mu m$.

The association of an RRL star to Trumpler~5, a decidedly intermediate-age simple stellar population, is the most direct evidence so far of the existence of these `young' RRLs and adds firm support to the evidence that has been systematically accumulating toward the existence of RRL stars at these unusual ages, both in the MW and LMC  \citep{Iorio2021,Sarbadhicary_2021,Zhang2025,CabreraGadea2025,CuevasOtahola2025}.

\begin{acknowledgements}
The authors would like to thank the anonymous referee for a constructive discussion and useful suggestions. CM is delighted to thank Pau Ramos and Danny Horta for useful discussions and early readings of the manuscript, as well as Giuliano Iorio, Alexey Bobrick, Valentina D'Orazi, Zdenek Prudil and Tatiana Muraveva from the Synergy Team for useful comments. JJD and BCO acknowledge support from the RDT funding from Universidad de la República, Uruguay. This work has made use of data from the European 
Space Agency (ESA) mission {\it Gaia} (\url{https://www.cosmos.esa.int/gaia}), processed 
by the {\it Gaia} Data Processing and Analysis Consortium (DPAC, \url{https://www.cosmos.esa.int/web/gaia/dpac/consortium}). Funding for the DPAC has 
been provided by national institutions, in particular the institutions participating 
in the {\it Gaia} Multilateral Agreement.\\
\\
{\it Software:} Astropy \citep{astropy2018},
    Matplotlib \citep{mpl},
    Numpy \citep{numpy},
    Jupyter \citep{jupyter2016}, 
    TOPCAT \citep{Topcat2005,Stilts2006},
    dustmaps \citep{Green2018}
\end{acknowledgements}



\bibliographystyle{aa}
\bibliography{refsdtd} 

@ARTICLE{Fitzpatrick1999,
   author = {{Fitzpatrick}, E.~L.},
    title = "{Correcting for the Effects of Interstellar Extinction}",
  journal = {\pasp},
   eprint = {arXiv:astro-ph/9809387},
     year = 1999,
    month = jan,
   volume = 111,
    pages = {63-75},
      doi = {10.1086/316293},
   adsurl = {http://adsabs.harvard.edu/abs/1999PASP..111...63F}
}

@ARTICLE{Indebetouw2005,
   author = {{Indebetouw}, R. and {Mathis}, J.~S. and {Babler}, B.~L. and
        {Meade}, M.~R. and {Watson}, C. and {Whitney}, B.~A. and {Wolff}, M.~J. and ^M
        {Wolfire}, M.~G. and {Cohen}, M. and {Bania}, T.~M. and {Benjamin}, R.~A. and
        {Clemens}, D.~P. and {Dickey}, J.~M. and {Jackson}, J.~M. and 
        {Kobulnicky}, H.~A. and {Marston}, A.~P. and {Mercer}, E.~P. and 
        {Stauffer}, J.~R. and {Stolovy}, S.~R. and {Churchwell}, E.},
    title = "{The Wavelength Dependence of Interstellar Extinction from 1.25 to 8.0 {$\mu$}m Using GLIMPSE Data}",
  journal = {\apj},
   eprint = {arXiv:astro-ph/0406403},
     year = 2005,
    month = feb,
   volume = 619,
    pages = {931-938},
      doi = {10.1086/426679},
   adsurl = {http://adsabs.harvard.edu/abs/2005ApJ...619..931I}
}

@ARTICLE{Choi2016,
       author = {{Choi}, Jieun and {Dotter}, Aaron and {Conroy}, Charlie and {Cantiello}, Matteo and {Paxton}, Bill and {Johnson}, Benjamin D.},
        title = "{Mesa Isochrones and Stellar Tracks (MIST). I. Solar-scaled Models}",
      journal = {\apj},
         year = 2016,
        month = jun,
       volume = {823},
       number = {2},
          eid = {102},
        pages = {102},
          doi = {10.3847/0004-637X/823/2/102},
archivePrefix = {arXiv},
       eprint = {1604.08592},
 primaryClass = {astro-ph.SR},
       adsurl = {https://ui.adsabs.harvard.edu/abs/2016ApJ...823..102C}
}

@ARTICLE{Dotter2016,
       author = {{Dotter}, Aaron},
        title = "{MESA Isochrones and Stellar Tracks (MIST) 0: Methods for the Construction of Stellar Isochrones}",
      journal = {\apjs},
         year = 2016,
        month = jan,
       volume = {222},
       number = {1},
          eid = {8},
        pages = {8},
          doi = {10.3847/0067-0049/222/1/8},
archivePrefix = {arXiv},
       eprint = {1601.05144},
 primaryClass = {astro-ph.SR},
       adsurl = {https://ui.adsabs.harvard.edu/abs/2016ApJS..222....8D}
}

@ARTICLE{Dias2021,
       author = {{Dias}, W.~S. and {Monteiro}, H. and {Moitinho}, A. and {L{\'e}pine}, J.~R.~D. and {Carraro}, G. and {Paunzen}, E. and {Alessi}, B. and {Villela}, L.},
        title = "{Updated parameters of 1743 open clusters based on Gaia DR2}",
      journal = {\mnras},
         year = 2021,
        month = jun,
       volume = {504},
       number = {1},
        pages = {356-371},
          doi = {10.1093/mnras/stab770},
archivePrefix = {arXiv},
       eprint = {2103.12829},
 primaryClass = {astro-ph.SR},
       adsurl = {https://ui.adsabs.harvard.edu/abs/2021MNRAS.504..356D}
}

@INPROCEEDINGS{Bastian2016,
       author = {{Bastian}, N.},
        title = "{Young Massive Clusters: Their Population Properties, Formation and Evolution, and Their Relation to the Ancient Globular Clusters}",
    booktitle = {EAS Publications Series},
         year = 2016,
       editor = {{Moraux}, Estelle and {Lebreton}, Yveline and {Charbonnel}, Corinne},
       series = {EAS Publications Series},
       volume = {80-81},
        month = nov,
        pages = {5-37},
          doi = {10.1051/eas/1680002},
archivePrefix = {arXiv},
       eprint = {1606.09468},
 primaryClass = {astro-ph.GA},
       adsurl = {https://ui.adsabs.harvard.edu/abs/2016EAS....80....5B}
}

@article{Rain2021_dyn,
doi = {10.3847/1538-3881/abc1ee},
url = {https://dx.doi.org/10.3847/1538-3881/abc1ee},
year = {2020},
month = {dec},
publisher = {The American Astronomical Society},
volume = {161},
number = {1},
pages = {37},
author = {Rain, M. J. and Carraro, G. and Ahumada, J. A. and Villanova, S. and Boffin, H. and Monaco, L.},
title = {The Blue Straggler Population of the Open Clusters Trumpler 5, Trumpler 20, and NGC 2477},
journal = {The Astronomical Journal}
}

@ARTICLE{Ozdemir2025,
       author = {{{\"O}zdemir}, S. and {Af{\c{s}}ar}, M. and {Sneden}, C. and {VandenBerg}, D.~A. and {Denissenkov}, P.~A. and {Milone}, A.~P. and {Bozkurt}, Z. and {Oh}, H. and {Sokal}, K. and {Mace}, G.~N. and {Jaffe}, D.~T.},
        title = "{High-resolution infrared spectroscopy of the dust-obscured metal-poor open cluster Trumpler 5}",
      journal = {\aap},
         year = 2025,
        month = jul,
       volume = {699},
          eid = {A208},
        pages = {A208},
          doi = {10.1051/0004-6361/202452561},
archivePrefix = {arXiv},
       eprint = {2505.09722},
 primaryClass = {astro-ph.SR},
       adsurl = {https://ui.adsabs.harvard.edu/abs/2025A&A...699A.208O}
}

@ARTICLE{Marconi2022,
       author = {{Marconi}, Marcella and {Molinaro}, Roberto and {Dall'Ora}, Massimo and {Ripepi}, Vincenzo and {Musella}, Ilaria and {Bono}, Giuseppe and {Braga}, Vittorio and {Di Criscienzo}, Marcella and {Fiorentino}, Giuliana and {Leccia}, Silvio and {Monelli}, Matteo},
        title = "{New Theoretical Period-Luminosity-Metallicity Relations for RR Lyrae in the Rubin-LSST Filters}",
      journal = {\apj},
         year = 2022,
        month = jul,
       volume = {934},
       number = {1},
          eid = {29},
        pages = {29},
          doi = {10.3847/1538-4357/ac78ee},
archivePrefix = {arXiv},
       eprint = {2206.06470},
 primaryClass = {astro-ph.SR},
       adsurl = {https://ui.adsabs.harvard.edu/abs/2022ApJ...934...29M}
}

@ARTICLE{Marconi2021,
       author = {{Marconi}, M. and {Molinaro}, R. and {Ripepi}, V. and {Leccia}, S. and {Musella}, I. and {De Somma}, G. and {Gatto}, M. and {Moretti}, M.~I.},
        title = "{A theoretical scenario for Galactic RR Lyrae in the Gaia data base: constraints on the parallax offset}",
      journal = {\mnras},
         year = 2021,
        month = jan,
       volume = {500},
       number = {4},
        pages = {5009-5023},
          doi = {10.1093/mnras/staa3558},
archivePrefix = {arXiv},
       eprint = {2011.06675},
 primaryClass = {astro-ph.SR},
       adsurl = {https://ui.adsabs.harvard.edu/abs/2021MNRAS.500.5009M}
}

@ARTICLE{PortegiesZwart2010,
       author = {{Portegies Zwart}, Simon F. and {McMillan}, Stephen L.~W. and {Gieles}, Mark},
        title = "{Young Massive Star Clusters}",
      journal = {\araa},
         year = 2010,
        month = sep,
       volume = {48},
        pages = {431-493},
          doi = {10.1146/annurev-astro-081309-130834},
archivePrefix = {arXiv},
       eprint = {1002.1961},
 primaryClass = {astro-ph.GA},
       adsurl = {https://ui.adsabs.harvard.edu/abs/2010ARA&A..48..431P}
}

@ARTICLE{Bayo2008,
       author = {{Bayo}, A. and {Rodrigo}, C. and {Barrado Y Navascu{\'e}s}, D. and {Solano}, E. and {Guti{\'e}rrez}, R. and {Morales-Calder{\'o}n}, M. and {Allard}, F.},
        title = "{VOSA: virtual observatory SED analyzer. An application to the Collinder 69 open cluster}",
      journal = {\aap},
         year = 2008,
        month = dec,
       volume = {492},
       number = {1},
        pages = {277-287},
          doi = {10.1051/0004-6361:200810395},
archivePrefix = {arXiv},
       eprint = {0808.0270},
 primaryClass = {astro-ph},
       adsurl = {https://ui.adsabs.harvard.edu/abs/2008A&A...492..277B}
}

@ARTICLE{Green2019,
       author = {{Green}, Gregory M. and {Schlafly}, Edward and {Zucker}, Catherine and {Speagle}, Joshua S. and {Finkbeiner}, Douglas},
        title = "{A 3D Dust Map Based on Gaia, Pan-STARRS 1, and 2MASS}",
      journal = {\apj},
         year = 2019,
        month = dec,
       volume = {887},
       number = {1},
          eid = {93},
        pages = {93},
          doi = {10.3847/1538-4357/ab5362},
archivePrefix = {arXiv},
       eprint = {1905.02734},
 primaryClass = {astro-ph.GA},
       adsurl = {https://ui.adsabs.harvard.edu/abs/2019ApJ...887...93G}
}

@article{Green2018, doi = {10.21105/joss.00695}, url = {https://doi.org/10.21105/joss.00695}, year = {2018}, publisher = {The Open Journal}, volume = {3}, number = {26}, pages = {695}, author = {Green, Gregory M.}, title = {dustmaps: A Python interface for maps of interstellar dust}, journal = {Journal of Open Source Software} }

@ARTICLE{Leiner2016,
       author = {{Leiner}, Emily and {Mathieu}, Robert D. and {Stello}, Dennis and {Vanderburg}, Andrew and {Sandquist}, Eric},
        title = "{The K2 M67 Study: An Evolved Blue Straggler in M67 from K2 Mission Asteroseismology}",
      journal = {\apjl},
         year = 2016,
        month = nov,
       volume = {832},
       number = {1},
          eid = {L13},
        pages = {L13},
          doi = {10.3847/2041-8205/832/1/L13},
archivePrefix = {arXiv},
       eprint = {1611.01158},
 primaryClass = {astro-ph.SR},
       adsurl = {https://ui.adsabs.harvard.edu/abs/2016ApJ...832L..13L}
}

@INPROCEEDINGS{Landsman1998,
       author = {{Landsman}, W. and {Simon}, T.},
        title = "{Helium White Dwarf Companions of Field Late-Type Stars}",
    booktitle = {American Astronomical Society Meeting Abstracts},
         year = 1998,
       series = {American Astronomical Society Meeting Abstracts},
       volume = {193},
        month = dec,
          eid = {37.03},
        pages = {37.03},
       adsurl = {https://ui.adsabs.harvard.edu/abs/1998AAS...193.3703L}
}

@ARTICLE{Ngeow2025,
       author = {{Ngeow}, Chow-Choong and {Bhardwaj}, Anupam},
        title = "{Candidate RR Lyrae Associated with the Ultrafaint Dwarf Galaxy Aquarius III}",
      journal = {\aj},
         year = 2025,
        month = mar,
       volume = {169},
       number = {3},
          eid = {156},
        pages = {156},
          doi = {10.3847/1538-3881/adadf1},
archivePrefix = {arXiv},
       eprint = {2502.03764},
 primaryClass = {astro-ph.GA},
       adsurl = {https://ui.adsabs.harvard.edu/abs/2025AJ....169..156N}
}

@ARTICLE{Lallement2022,
       author = {{Lallement}, R. and {Vergely}, J.~L. and {Babusiaux}, C. and {Cox}, N.~L.~J.},
        title = "{Updated Gaia-2MASS 3D maps of Galactic interstellar dust}",
      journal = {\aap},
         year = 2022,
        month = may,
       volume = {661},
          eid = {A147},
        pages = {A147},
          doi = {10.1051/0004-6361/202142846},
archivePrefix = {arXiv},
       eprint = {2203.01627},
 primaryClass = {astro-ph.GA},
       adsurl = {https://ui.adsabs.harvard.edu/abs/2022A&A...661A.147L}
}

@ARTICLE{Marconi2015,
       author = {{Marconi}, M. and {Coppola}, G. and {Bono}, G. and {Braga}, V. and {Pietrinferni}, A. and {Buonanno}, R. and {Castellani}, M. and {Musella}, I. and {Ripepi}, V. and {Stellingwerf}, R.~F.},
        title = "{On a New Theoretical Framework for RR Lyrae Stars. I. The Metallicity Dependence}",
      journal = {\apj},
         year = 2015,
        month = jul,
       volume = {808},
       number = {1},
          eid = {50},
        pages = {50},
          doi = {10.1088/0004-637X/808/1/50},
archivePrefix = {arXiv},
       eprint = {1505.02531},
 primaryClass = {astro-ph.SR},
       adsurl = {https://ui.adsabs.harvard.edu/abs/2015ApJ...808...50M}
}

@ARTICLE{Mucciarelli2021,
       author = {{Mucciarelli}, A. and {Bellazzini}, M. and {Massari}, D.},
        title = "{Exploiting the Gaia EDR3 photometry to derive stellar temperatures}",
      journal = {\aap},
         year = 2021,
        month = sep,
       volume = {653},
          eid = {A90},
        pages = {A90},
          doi = {10.1051/0004-6361/202140979},
archivePrefix = {arXiv},
       eprint = {2106.03882},
 primaryClass = {astro-ph.SR},
       adsurl = {https://ui.adsabs.harvard.edu/abs/2021A&A...653A..90M}
}

@ARTICLE{PecautMamajek2013,
       author = {{Pecaut}, Mark J. and {Mamajek}, Eric E.},
        title = "{Intrinsic Colors, Temperatures, and Bolometric Corrections of Pre-main-sequence Stars}",
      journal = {\apjs},
         year = 2013,
        month = sep,
       volume = {208},
       number = {1},
          eid = {9},
        pages = {9},
          doi = {10.1088/0067-0049/208/1/9},
archivePrefix = {arXiv},
       eprint = {1307.2657},
 primaryClass = {astro-ph.SR},
       adsurl = {https://ui.adsabs.harvard.edu/abs/2013ApJS..208....9P}
}

@ARTICLE{Fitzpatrick2019,
       author = {{Fitzpatrick}, E.~L. and {Massa}, Derck and {Gordon}, Karl D. and {Bohlin}, Ralph and {Clayton}, Geoffrey C.},
        title = "{An Analysis of the Shapes of Interstellar Extinction Curves. VII. Milky Way Spectrophotometric Optical-through-ultraviolet Extinction and Its R-dependence}",
      journal = {\apj},
         year = 2019,
        month = dec,
       volume = {886},
       number = {2},
          eid = {108},
        pages = {108},
          doi = {10.3847/1538-4357/ab4c3a},
archivePrefix = {arXiv},
       eprint = {1910.08852},
 primaryClass = {astro-ph.GA},
       adsurl = {https://ui.adsabs.harvard.edu/abs/2019ApJ...886..108F}
}

@ARTICLE{DOrazi2024,
       author = {{D'Orazi}, Valentina and {Storm}, Nicholas and {Casey}, Andrew R. and {Braga}, Vittorio F. and {Zocchi}, Alice and {Bono}, Giuseppe and {Fabrizio}, Michele and {Sneden}, Christopher and {Massari}, Davide and {Giribaldi}, Riano E. and {Bergemann}, Maria and {Campbell}, Simon W. and {Casagrande}, Luca and {de Grijs}, Richard and {De Silva}, Gayandhi and {Lugaro}, Maria and {Zucker}, Daniel B. and {Bragaglia}, Angela and {Feuillet}, Diane and {Fiorentino}, Giuliana and {Chaboyer}, Brian and {Dall'Ora}, Massimo and {Marengo}, Massimo and {Mart{\'\i}nez-V{\'a}zquez}, Clara E. and {Matsunaga}, Noriyuki and {Monelli}, Matteo and {Mullen}, Joseph P. and {Nataf}, David and {Tantalo}, Maria and {Thevenin}, Frederic and {Vitello}, Fabio R. and {Kudritzki}, Rolf-Peter and {Bland-Hawthorn}, Joss and {Buder}, Sven and {Freeman}, Ken and {Kos}, Janez and {Lewis}, Geraint F. and {Lind}, Karin and {Martell}, Sarah and {Sharma}, Sanjib and {Stello}, Dennis and {Zwitter}, Toma{\v{z}}},
        title = "{The GALAH survey: tracing the Milky Way's formation and evolution through RR Lyrae stars}",
      journal = {\mnras},
         year = 2024,
        month = jun,
       volume = {531},
       number = {1},
        pages = {137-162},
          doi = {10.1093/mnras/stae1149},
archivePrefix = {arXiv},
       eprint = {2405.04580},
 primaryClass = {astro-ph.GA},
       adsurl = {https://ui.adsabs.harvard.edu/abs/2024MNRAS.531..137D}
}

@ARTICLE{Prudil2025,
       author = {{Prudil}, Z. and {Debattista}, V.~P. and {Beraldo e Silva}, L. and {Anderson}, S.~R. and {Gough-Kelly}, S. and {Kunder}, A. and {Rejkuba}, M. and {Gerhard}, O. and {Wyse}, R.~F.~G. and {Koch-Hansen}, A.~J. and {Rich}, R.~M. and {Savino}, A.},
        title = "{The Galactic bulge exploration: V. The secular spherical and X-shaped Milky Way bulge}",
      journal = {\aap},
         year = 2025,
        month = jul,
       volume = {699},
          eid = {A349},
        pages = {A349},
          doi = {10.1051/0004-6361/202554819},
archivePrefix = {arXiv},
       eprint = {2506.19074},
 primaryClass = {astro-ph.GA},
       adsurl = {https://ui.adsabs.harvard.edu/abs/2025A&A...699A.349P}
}

@INPROCEEDINGS{Landsman1997,
       author = {{Landsman}, Wayne and {Stecher}, Theodore P.},
        title = "{Ultraviolet imagery of the old open clusters M67, NGC 188, and NGC 6791}",
    booktitle = {The ultraviolet universe at low and High redshift},
         year = 1997,
       editor = {{Waller}, William H.},
       series = {American Institute of Physics Conference Series},
       volume = {408},
        month = may,
        pages = {390-394},
          doi = {10.1063/1.53761},
       adsurl = {https://ui.adsabs.harvard.edu/abs/1997AIPC..408..390L}
}

@ARTICLE{Donati2015,
       author = {{Donati}, P. and {Cocozza}, G. and {Bragaglia}, A. and {Pancino}, E. and {Cantat-Gaudin}, T. and {Carrera}, R. and {Tosi}, M.},
        title = "{The old, metal-poor, anticentre open cluster Trumpler 5}",
      journal = {\mnras},
         year = 2015,
        month = jan,
       volume = {446},
       number = {2},
        pages = {1411-1423},
          doi = {10.1093/mnras/stu2162},
archivePrefix = {arXiv},
       eprint = {1411.0717},
 primaryClass = {astro-ph.SR},
       adsurl = {https://ui.adsabs.harvard.edu/abs/2015MNRAS.446.1411D}
}

@ARTICLE{Kinman2010,
       author = {{Kinman}, T.~D. and {Brown}, Warren R.},
        title = "{Low-amplitude Variables: Distinguishing RR Lyrae Stars from Eclipsing Binaries}",
      journal = {\aj},
         year = 2010,
        month = may,
       volume = {139},
       number = {5},
        pages = {2014-2025},
          doi = {10.1088/0004-6256/139/5/2014},
archivePrefix = {arXiv},
       eprint = {1003.3656},
 primaryClass = {astro-ph.GA},
       adsurl = {https://ui.adsabs.harvard.edu/abs/2010AJ....139.2014K}
}

@ARTICLE{Rain2021,
       author = {{Rain}, M.~J. and {Ahumada}, J.~A. and {Carraro}, G.},
        title = "{A new, Gaia-based, catalogue of blue straggler stars in open clusters}",
      journal = {\aap},
         year = 2021,
        month = jun,
       volume = {650},
          eid = {A67},
        pages = {A67},
          doi = {10.1051/0004-6361/202040072},
archivePrefix = {arXiv},
       eprint = {2103.06004},
 primaryClass = {astro-ph.SR},
       adsurl = {https://ui.adsabs.harvard.edu/abs/2021A&A...650A..67R}
}

@ARTICLE{Mermilliod2007,
       author = {{Mermilliod}, J. -C. and {Andersen}, J. and {Latham}, D.~W. and {Mayor}, M.},
        title = "{Red giants in open clusters. XIII. Orbital elements of 156 spectroscopic binaries}",
      journal = {\aap},
         year = 2007,
        month = oct,
       volume = {473},
       number = {3},
        pages = {829-845},
          doi = {10.1051/0004-6361:20078007},
       adsurl = {https://ui.adsabs.harvard.edu/abs/2007A&A...473..829M}
}

@ARTICLE{Anderson2025,
       author = {{Anderson}, Richard I. and {Hunt}, Emily L.},
        title = "{A bird's eye view of stellar evolution through populations of variable stars in Galactic open clusters}",
      journal = {\aap},
         year = 2025,
        month = aug,
       volume = {700},
          eid = {L13},
        pages = {L13},
          doi = {10.1051/0004-6361/202555111},
archivePrefix = {arXiv},
       eprint = {2508.12866},
 primaryClass = {astro-ph.SR},
       adsurl = {https://ui.adsabs.harvard.edu/abs/2025A&A...700L..13A}
}

@ARTICLE{CabreraGadea2025,
       author = {{Cabrera-Gadea}, Mauro and {Mateu}, Cecilia and {Ramos}, Pau},
        title = "{RR Lyrae stars trace the Milky Way warp}",
      journal = {\aap},
         year = 2025,
        month = sep,
       volume = {701},
          eid = {A136},
        pages = {A136},
          doi = {10.1051/0004-6361/202452736},
archivePrefix = {arXiv},
       eprint = {2410.22427},
 primaryClass = {astro-ph.GA},
       adsurl = {https://ui.adsabs.harvard.edu/abs/2025A&A...701A.136C}
}

@ARTICLE{Hunt2024,
       author = {{Hunt}, Emily L. and {Reffert}, Sabine},
        title = "{Improving the open cluster census. III. Using cluster masses, radii, and dynamics to create a cleaned open cluster catalogue}",
      journal = {\aap},
         year = 2024,
        month = jun,
       volume = {686},
          eid = {A42},
        pages = {A42},
          doi = {10.1051/0004-6361/202348662},
archivePrefix = {arXiv},
       eprint = {2403.05143},
 primaryClass = {astro-ph.GA},
       adsurl = {https://ui.adsabs.harvard.edu/abs/2024A&A...686A..42H}
}

@ARTICLE{Hunt2023,
       author = {{Hunt}, Emily L. and {Reffert}, Sabine},
        title = "{Improving the open cluster census. II. An all-sky cluster catalogue with Gaia DR3}",
      journal = {\aap},
         year = 2023,
        month = may,
       volume = {673},
          eid = {A114},
        pages = {A114},
          doi = {10.1051/0004-6361/202346285},
archivePrefix = {arXiv},
       eprint = {2303.13424},
 primaryClass = {astro-ph.GA},
       adsurl = {https://ui.adsabs.harvard.edu/abs/2023A&A...673A.114H}
}

@ARTICLE{vanGroeningen2023,
       author = {{van Groeningen}, M.~G.~J. and {Castro-Ginard}, A. and {Brown}, A.~G.~A. and {Casamiquela}, L. and {Jordi}, C.},
        title = "{A machine-learning-based tool for open cluster membership determination in Gaia DR3}",
      journal = {\aap},
         year = 2023,
        month = jul,
       volume = {675},
          eid = {A68},
        pages = {A68},
          doi = {10.1051/0004-6361/202345952},
archivePrefix = {arXiv},
       eprint = {2303.08474},
 primaryClass = {astro-ph.GA},
       adsurl = {https://ui.adsabs.harvard.edu/abs/2023A&A...675A..68V}
}

@ARTICLE{CantatGaudin2018,
       author = {{Cantat-Gaudin}, T. and {Jordi}, C. and {Vallenari}, A. and {Bragaglia}, A. and {Balaguer-N{\'u}{\~n}ez}, L. and {Soubiran}, C. and {Bossini}, D. and {Moitinho}, A. and {Castro-Ginard}, A. and {Krone-Martins}, A. and {Casamiquela}, L. and {Sordo}, R. and {Carrera}, R.},
        title = "{A Gaia DR2 view of the open cluster population in the Milky Way}",
      journal = {\aap},
         year = 2018,
        month = oct,
       volume = {618},
          eid = {A93},
        pages = {A93},
          doi = {10.1051/0004-6361/201833476},
archivePrefix = {arXiv},
       eprint = {1805.08726},
 primaryClass = {astro-ph.GA},
       adsurl = {https://ui.adsabs.harvard.edu/abs/2018A&A...618A..93C}
}

@ARTICLE{Ferraro2026,
       author = {{Ferraro}, Francesco R. and {Lanzoni}, Barbara and {Vesperini}, Enrico and {Dalessandro}, Emanuele and {Cadelano}, Mario and {Pallanca}, Cristina and {Beccari}, Giacomo and {Nardiello}, Domenico and {Libralato}, Mattia and {Piotto}, Giampaolo},
        title = "{A binary-related origin mediated by environmental conditions for blue straggler stars}",
      journal = {Nature Communications},
         year = 2026,
        month = jan,
       volume = {17},
       number = {1},
          eid = {768},
        pages = {768},
          doi = {10.1038/s41467-025-68159-5},
archivePrefix = {arXiv},
       eprint = {2506.07692},
 primaryClass = {astro-ph.SR},
       adsurl = {https://ui.adsabs.harvard.edu/abs/2026NatCo..17..768F}
}

@INPROCEEDINGS{Momany2015,
       author = {{Momany}, Yazan},
        title = "{The Blue Straggler Population in Dwarf Galaxies}",
    booktitle = {Astrophysics and Space Science Library},
         year = 2015,
       editor = {{Boffin}, Henri M.~J. and {Carraro}, Giovanni and {Beccari}, Giacomo},
       series = {Astrophysics and Space Science Library},
       volume = {413},
        month = jan,
        pages = {129},
          doi = {10.1007/978-3-662-44434-4_6},
archivePrefix = {arXiv},
       eprint = {1406.3472},
 primaryClass = {astro-ph.SR},
       adsurl = {https://ui.adsabs.harvard.edu/abs/2015ASSL..413..129M}
}

@ARTICLE{CantatGaudin2020,
       author = {{Cantat-Gaudin}, T. and {Anders}, F. and {Castro-Ginard}, A. and {Jordi}, C. and {Romero-G{\'o}mez}, M. and {Soubiran}, C. and {Casamiquela}, L. and {Tarricq}, Y. and {Moitinho}, A. and {Vallenari}, A. and {Bragaglia}, A. and {Krone-Martins}, A. and {Kounkel}, M.},
        title = "{Painting a portrait of the Galactic disc with its stellar clusters}",
      journal = {\aap},
         year = 2020,
        month = aug,
       volume = {640},
          eid = {A1},
        pages = {A1},
          doi = {10.1051/0004-6361/202038192},
archivePrefix = {arXiv},
       eprint = {2004.07274},
 primaryClass = {astro-ph.GA},
       adsurl = {https://ui.adsabs.harvard.edu/abs/2020A&A...640A...1C}
}

@ARTICLE{CuevasOtahola2025,
       author = {{Cuevas-Otahola}, Bolivia and {Mateu}, Cecilia and {Cabrera-Ziri}, Ivan and {Bruzual}, Gustavo and {Hern{\'a}ndez-P{\'e}rez}, Fabiola and {Magris}, Gladis and {Baumgardt}, Holger},
        title = "{RR Lyrae stars in intermediate-age Magellanic clusters: membership probabilities and delay time distribution}",
      journal = {\mnras},
         year = 2025,
        month = aug,
       volume = {541},
       number = {2},
        pages = {1434-1448},
          doi = {10.1093/mnras/staf1095},
archivePrefix = {arXiv},
       eprint = {2411.12741},
 primaryClass = {astro-ph.GA},
       adsurl = {https://ui.adsabs.harvard.edu/abs/2025MNRAS.541.1434C}
}

@ARTICLE{Medina2024,
       author = {{Medina}, Gustavo E. and {Mu{\~n}oz}, Ricardo R. and {Carlin}, Jeffrey L. and {Vivas}, A. Katherina and {Grebel}, Eva K. and {Mart{\'\i}nez-V{\'a}zquez}, Clara E. and {Hansen}, Camilla J.},
        title = "{Taking the pulse of the outer Milky Way with the Halo Outskirts With Variable Stars (HOWVAST) survey: an RR Lyrae density profile out to >200 kpc}",
      journal = {\mnras},
         year = 2024,
        month = jul,
       volume = {531},
       number = {4},
        pages = {4762-4780},
          doi = {10.1093/mnras/stae1137},
archivePrefix = {arXiv},
       eprint = {2402.14055},
 primaryClass = {astro-ph.GA},
       adsurl = {https://ui.adsabs.harvard.edu/abs/2024MNRAS.531.4762M}
}

@ARTICLE{Li2023,
       author = {{Li}, Xin-Yi and {Huang}, Yang and {Liu}, Gao-Chao and {Beers}, Timothy C. and {Zhang}, Hua-Wei},
        title = "{Photometric Metallicity and Distance Estimates for 136,000 RR Lyrae Stars from Gaia Data Release 3}",
      journal = {\apj},
         year = 2023,
        month = feb,
       volume = {944},
       number = {1},
          eid = {88},
        pages = {88},
          doi = {10.3847/1538-4357/acadd5},
archivePrefix = {arXiv},
       eprint = {2206.07668},
 primaryClass = {astro-ph.SR},
       adsurl = {https://ui.adsabs.harvard.edu/abs/2023ApJ...944...88L}
}

@ARTICLE{Hajdu2021,
       author = {{Hajdu}, Gergely and {Pietrzy{\'n}ski}, Grzegorz and {Jurcsik}, Johanna and {Catelan}, M{\'a}rcio and {Karczmarek}, Paulina and {Pilecki}, Bogumi{\l} and {Soszy{\'n}ski}, Igor and {Udalski}, Andrzej and {Thompson}, Ian B.},
        title = "{Studies of RR Lyrae Variables in Binary Systems. I. Evidence of a Trimodal Companion Mass Distribution}",
      journal = {\apj},
         year = 2021,
        month = jul,
       volume = {915},
       number = {1},
          eid = {50},
        pages = {50},
          doi = {10.3847/1538-4357/abff4b},
archivePrefix = {arXiv},
       eprint = {2105.03750},
 primaryClass = {astro-ph.SR},
       adsurl = {https://ui.adsabs.harvard.edu/abs/2021ApJ...915...50H}
}

@article{Smith1995,
author = {Horace A Smith},
journal = {RR Lyrae Stars, Cambridge Astrophysics Series},
title = {RR Lyrae stars},
year = {1995},
month = {Jan},
pmid = {1995rls..book.....S},
URL = {http://adsabs.harvard.edu/cgi-bin/nph-data_query?bibcode=1995rls..book.....S&link_type=CITATIONS},
uri = {papers://3CF19FFD-9F37-4577-AFDC-1D7F0F338DCB/Paper/p3940}
}

@book{CatelanSmith2015,
	address = {Weinheim, Bergstr},
	edition = {1. Auflage},
	title = {Pulsating {Stars}},
	isbn = {978-3-527-40715-6},
	language = {eng},
	publisher = {Wiley-VCH},
	author = {Catelan, Márcio and Smith, Horace A.},
	year = {2015},
}

@ARTICLE{Prudil2020,
       author = {{Prudil}, Z. and {D{\'e}k{\'a}ny}, I. and {Grebel}, E.~K. and {Kunder}, A.},
        title = "{Evidence for Galactic disc RR Lyrae stars in the solar neighbourhood}",
      journal = {\mnras},
         year = 2020,
        month = mar,
       volume = {492},
       number = {3},
        pages = {3408-3419},
          doi = {10.1093/mnras/staa046},
archivePrefix = {arXiv},
       eprint = {2001.02486},
 primaryClass = {astro-ph.SR},
       adsurl = {https://ui.adsabs.harvard.edu/abs/2020MNRAS.492.3408P}
}

@ARTICLE{Taam1976,
       author = {{Taam}, R.~E. and {Kraft}, R.~P. and {Suntzeff}, N.},
        title = "{The origin and evolution of RR Lyrae stars of high metal abundance.}",
      journal = {\apj},
         year = 1976,
        month = jul,
       volume = {207},
        pages = {201-208},
          doi = {10.1086/154485},
       adsurl = {https://ui.adsabs.harvard.edu/abs/1976ApJ...207..201T}
}

@ARTICLE{Bono1997b,
       author = {{Bono}, Giuseppe and {Caputo}, Filippina and {Cassisi}, Santi and {Incerpi}, Roberta and {Marconi}, Marcella},
        title = "{Metal-rich RR Lyrae Variables. II. The Pulsational Scenario}",
      journal = {\apj},
         year = 1997,
        month = jul,
       volume = {483},
       number = {2},
        pages = {811-825},
          doi = {10.1086/304284},
archivePrefix = {arXiv},
       eprint = {astro-ph/9702083},
 primaryClass = {astro-ph},
       adsurl = {https://ui.adsabs.harvard.edu/abs/1997ApJ...483..811B}
}

@ARTICLE{Bono1997a,
       author = {{Bono}, Giuseppe and {Caputo}, Filippina and {Cassisi}, Santi and {Castellani}, Vittorio and {Marconi}, Marcella},
        title = "{Metal-rich RR Lyrae Variables. I. The Evolutionary Scenario}",
      journal = {\apj},
         year = 1997,
        month = apr,
       volume = {479},
       number = {1},
        pages = {279-289},
          doi = {10.1086/303872},
archivePrefix = {arXiv},
       eprint = {astro-ph/9609153},
 primaryClass = {astro-ph},
       adsurl = {https://ui.adsabs.harvard.edu/abs/1997ApJ...479..279B}
}

@UNPUBLISHED{Lindegren2018,
   author = {L.~Lindegren},
   title={{R}e-normalising the astrometric chi-square in {G}aia {D}{R}2},
   institution={Lund Observatory},
   year={2018},
   month={August},
   url={http://www.rssd.esa.int/doc_fetch.php?id=3757412},
   note={GAIA-C3-TN-LU-LL-124},
   type={Technical Note}
}

@ARTICLE{He2025,
       author = {{He}, Shun-Xuan and {Huang}, Yang and {Li}, Xin-Yi and {Zhang}, Hua-Wei and {Liu}, Gao-Chao and {Beers}, Timothy C. and {Wu}, Hong and {Fan}, Zhou},
        title = "{Photometric Metallicity and Distance Estimates for {\ensuremath{\sim}}70,000 RR Lyrae Stars from the Zwicky Transient Facility}",
      journal = {\apjs},
         year = 2025,
        month = may,
       volume = {278},
       number = {1},
          eid = {2},
        pages = {2},
          doi = {10.3847/1538-4365/adbcad},
archivePrefix = {arXiv},
       eprint = {2503.03378},
 primaryClass = {astro-ph.SR},
       adsurl = {https://ui.adsabs.harvard.edu/abs/2025ApJS..278....2H}
}

@ARTICLE{Muraveva2025,
       author = {{Muraveva}, Tatiana and {Giannetti}, Andrea and {Clementini}, Gisella and {Garofalo}, Alessia and {Monti}, Lorenzo},
        title = "{Metallicity of RR Lyrae stars from the Gaia Data Release 3 catalogue computed with Machine Learning algorithms}",
      journal = {\mnras},
         year = 2025,
        month = jan,
       volume = {536},
       number = {3},
        pages = {2749-2769},
          doi = {10.1093/mnras/stae2679},
archivePrefix = {arXiv},
       eprint = {2407.05815},
 primaryClass = {astro-ph.SR},
       adsurl = {https://ui.adsabs.harvard.edu/abs/2025MNRAS.536.2749M}
}

@ARTICLE{Zhang2025,
       author = {{Zh{\={a}}ng}, HanYuan and {Iorio}, Giuliano and {Belokurov}, Vasily and {Evans}, N. Wyn and {Bobrick}, Alexey and {D'Orazi}, Valentina},
        title = "{Revealing the ages of metal-rich RR Lyrae via kinematic label transfer}",
      journal = {\mnras},
         year = 2025,
        month = dec,
       volume = {544},
       number = {2},
        pages = {2493-2512},
          doi = {10.1093/mnras/staf1789},
archivePrefix = {arXiv},
       eprint = {2504.06720},
 primaryClass = {astro-ph.SR},
       adsurl = {https://ui.adsabs.harvard.edu/abs/2025MNRAS.544.2493Z}
}

@ARTICLE{Karczmarek2017,
       author = {{Karczmarek}, P. and {Wiktorowicz}, G. and {I{\l}kiewicz}, K. and {Smolec}, R. and {St{\k{e}}pie{\'n}}, K. and {Pietrzy{\'n}ski}, G. and {Gieren}, W. and {Belczynski}, K.},
        title = "{The occurrence of binary evolution pulsators in classical instability strip of RR Lyrae and Cepheid variables}",
      journal = {\mnras},
         year = 2017,
        month = apr,
       volume = {466},
       number = {3},
        pages = {2842-2854},
          doi = {10.1093/mnras/stw3286},
archivePrefix = {arXiv},
       eprint = {1612.00465},
 primaryClass = {astro-ph.SR},
       adsurl = {https://ui.adsabs.harvard.edu/abs/2017MNRAS.466.2842K}
}

@ARTICLE{Chen2020,
       author = {{Chen}, Xiaodian and {Wang}, Shu and {Deng}, Licai and {de Grijs}, Richard and {Yang}, Ming and {Tian}, Hao},
        title = "{The Zwicky Transient Facility Catalog of Periodic Variable Stars}",
      journal = {\apjs},
         year = 2020,
        month = jul,
       volume = {249},
       number = {1},
          eid = {18},
        pages = {18},
          doi = {10.3847/1538-4365/ab9cae},
archivePrefix = {arXiv},
       eprint = {2005.08662},
 primaryClass = {astro-ph.SR},
       adsurl = {https://ui.adsabs.harvard.edu/abs/2020ApJS..249...18C}
}

@ARTICLE{Jayasinghe2019a,
       author = {{Jayasinghe}, T. and {Stanek}, K.~Z. and {Kochanek}, C.~S. and {Shappee}, B.~J. and {Holoien}, T.~W. -S. and {Thompson}, Todd A. and {Prieto}, J.~L. and {Dong}, Subo and {Pawlak}, M. and {Pejcha}, O. and {Shields}, J.~V. and {Pojmanski}, G. and {Otero}, S. and {Britt}, C.~A. and {Will}, D.},
        title = "{The ASAS-SN catalogue of variable stars - II. Uniform classification of 412 000 known variables}",
      journal = {\mnras},
         year = 2019,
        month = jun,
       volume = {486},
       number = {2},
        pages = {1907-1943},
          doi = {10.1093/mnras/stz844},
archivePrefix = {arXiv},
       eprint = {1809.07329},
 primaryClass = {astro-ph.SR},
       adsurl = {https://ui.adsabs.harvard.edu/abs/2019MNRAS.486.1907J}
}

@ARTICLE{Jayasinghe2019b,
       author = {{Jayasinghe}, T. and {Stanek}, K.~Z. and {Kochanek}, C.~S. and {Shappee}, B.~J. and {Holoien}, T.~W. -S. and {Thompson}, Todd A. and {Prieto}, J.~L. and {Dong}, Subo and {Pawlak}, M. and {Pejcha}, O. and {Shields}, J.~V. and {Pojmanski}, G. and {Otero}, S. and {Hurst}, N. and {Britt}, C.~A. and {Will}, D.},
        title = "{The ASAS-SN catalogue of variable stars III: variables in the southern TESS continuous viewing zone}",
      journal = {\mnras},
         year = 2019,
        month = may,
       volume = {485},
       number = {1},
        pages = {961-971},
          doi = {10.1093/mnras/stz444},
archivePrefix = {arXiv},
       eprint = {1901.00009},
 primaryClass = {astro-ph.SR},
       adsurl = {https://ui.adsabs.harvard.edu/abs/2019MNRAS.485..961J}
}

@ARTICLE{Soszynski2016,
       author = {{Soszy{\'n}ski}, I. and {Udalski}, A. and {Szyma{\'n}ski}, M.~K. and {Wyrzykowski}, {\L}. and {Ulaczyk}, K. and {Poleski}, R. and {Pietrukowicz}, P. and {Koz{\l}owski}, S. and {Skowron}, D.~M. and {Skowron}, J. and {Mr{\'o}z}, P. and {Pawlak}, M.},
        title = "{The OGLE Collection of Variable Stars. Over 45 000 RR Lyrae Stars in the Magellanic System}",
      journal = {\actaa},
         year = 2016,
        month = jun,
       volume = {66},
       number = {2},
        pages = {131-147},
archivePrefix = {arXiv},
       eprint = {1606.02727},
 primaryClass = {astro-ph.SR},
       adsurl = {https://ui.adsabs.harvard.edu/abs/2016AcA....66..131S}
}

@ARTICLE{Sesar2017,
       author = {{Sesar}, Branimir and {Hernitschek}, Nina and {Dierickx}, Marion I.~P. and {Fardal}, Mark A. and {Rix}, Hans-Walter},
        title = "{The >100 kpc Distant Spur of the Sagittarius Stream and the Outer Virgo Overdensity, as Seen in PS1 RR Lyrae Stars}",
      journal = {\apjl},
         year = 2017,
        month = jul,
       volume = {844},
       number = {1},
          eid = {L4},
        pages = {L4},
          doi = {10.3847/2041-8213/aa7c61},
archivePrefix = {arXiv},
       eprint = {1706.10187},
 primaryClass = {astro-ph.GA},
       adsurl = {https://ui.adsabs.harvard.edu/abs/2017ApJ...844L...4S}
}

@ARTICLE{Sesar2017c,
   author = {{Sesar}, B. and {Hernitschek}, N. and {Mitrovi{\'c}}, S. and 
	{Ivezi{\'c}}, {\v Z}. and {Rix}, H.-W. and {Cohen}, J.~G. and 
	{Bernard}, E.~J. and {Grebel}, E.~K. and {Martin}, N.~F. and 
	{Schlafly}, E.~F. and {Burgett}, W.~S. and {Draper}, P.~W. and 
	{Flewelling}, H. and {Kaiser}, N. and {Kudritzki}, R.~P. and 
	{Magnier}, E.~A. and {Metcalfe}, N. and {Tonry}, J.~L. and {Waters}, C.
	},
    title = "{Machine-learned Identification of RR Lyrae Stars from Sparse, Multi-band Data: The PS1 Sample}",
  journal = {\aj},
archivePrefix = "arXiv",
   eprint = {1611.08596},
     year = 2017,
    month = may,
   volume = 153,
      eid = {204},
    pages = {204},
      doi = {10.3847/1538-3881/aa661b},
   adsurl = {http://adsabs.harvard.edu/abs/2017AJ....153..204S}
}

@ARTICLE{Bobrick2024,
       author = {{Bobrick}, Alexey and {Iorio}, Giuliano and {Belokurov}, Vasily and {Vos}, Joris and {Vu{\v{c}}kovi{\'c}}, Maja and {Giacobbo}, Nicola},
        title = "{RR Lyrae from binary evolution: abundant, young, and metal-rich}",
      journal = {\mnras},
         year = 2024,
        month = feb,
       volume = {527},
       number = {4},
        pages = {12196-12218},
          doi = {10.1093/mnras/stad3996},
archivePrefix = {arXiv},
       eprint = {2208.04332},
 primaryClass = {astro-ph.SR},
       adsurl = {https://ui.adsabs.harvard.edu/abs/2024MNRAS.52712196B}
}

@ARTICLE{Mateu2018,
       author = {{Mateu}, Cecilia and {Vivas}, A. Katherina},
        title = "{The Galactic thick disc density profile traced with RR Lyrae stars}",
      journal = {\mnras},
         year = 2018,
        month = sep,
       volume = {479},
       number = {1},
        pages = {211-227},
          doi = {10.1093/mnras/sty1373},
archivePrefix = {arXiv},
       eprint = {1802.07798},
 primaryClass = {astro-ph.GA},
       adsurl = {https://ui.adsabs.harvard.edu/abs/2018MNRAS.479..211M}
}

@ARTICLE{GaiaCol_DR3,
       author = {{Gaia Collaboration} and {Vallenari}, A. and {Brown}, A.~G.~A. and {Prusti}, T. and {de Bruijne}, J.~H.~J. and {Arenou}, F. and {Babusiaux}, C. and {Biermann}, M. and {Creevey}, O.~L. and {Ducourant}, C. and {Evans}, D.~W. and {Eyer}, L. and {Guerra}, R. and {Hutton}, A. and {Jordi}, C. and {Klioner}, S.~A. and {Lammers}, U.~L. and {Lindegren}, L. and {Luri}, X. and {Mignard}, F. and {Panem}, C. and {Pourbaix}, D. and {Randich}, S. and {Sartoretti}, P. and {Soubiran}, C. and {Tanga}, P. and {Walton}, N.~A. and {Bailer-Jones}, C.~A.~L. and {Bastian}, U. and {Drimmel}, R. and {Jansen}, F. and {Katz}, D. and {Lattanzi}, M.~G. and {van Leeuwen}, F. and {Bakker}, J. and {Cacciari}, C. and {Casta{\~n}eda}, J. and {De Angeli}, F. and {Fabricius}, C. and {Fouesneau}, M. and {Fr{\'e}mat}, Y. and {Galluccio}, L. and {Guerrier}, A. and {Heiter}, U. and {Masana}, E. and {Messineo}, R. and {Mowlavi}, N. and {Nicolas}, C. and {Nienartowicz}, K. and {Pailler}, F. and {Panuzzo}, P. and {Riclet}, F. and {Roux}, W. and {Seabroke}, G.~M. and {Sordo}, R. and {Th{\'e}venin}, F. and {Gracia-Abril}, G. and {Portell}, J. and {Teyssier}, D. and {Altmann}, M. and {Andrae}, R. and {Audard}, M. and {Bellas-Velidis}, I. and {Benson}, K. and {Berthier}, J. and {Blomme}, R. and {Burgess}, P.~W. and {Busonero}, D. and {Busso}, G. and {C{\'a}novas}, H. and {Carry}, B. and {Cellino}, A. and {Cheek}, N. and {Clementini}, G. and {Damerdji}, Y. and {Davidson}, M. and {de Teodoro}, P. and {Nu{\~n}ez Campos}, M. and {Delchambre}, L. and {Dell'Oro}, A. and {Esquej}, P. and {Fern{\'a}ndez-Hern{\'a}ndez}, J. and {Fraile}, E. and {Garabato}, D. and {Garc{\'\i}a-Lario}, P. and {Gosset}, E. and {Haigron}, R. and {Halbwachs}, J. -L. and {Hambly}, N.~C. and {Harrison}, D.~L. and {Hern{\'a}ndez}, J. and {Hestroffer}, D. and {Hodgkin}, S.~T. and {Holl}, B. and {Jan{\ss}en}, K. and {Jevardat de Fombelle}, G. and {Jordan}, S. and {Krone-Martins}, A. and {Lanzafame}, A.~C. and {L{\"o}ffler}, W. and {Marchal}, O. and {Marrese}, P.~M. and {Moitinho}, A. and {Muinonen}, K. and {Osborne}, P. and {Pancino}, E. and {Pauwels}, T. and {Recio-Blanco}, A. and {Reyl{\'e}}, C. and {Riello}, M. and {Rimoldini}, L. and {Roegiers}, T. and {Rybizki}, J. and {Sarro}, L.~M. and {Siopis}, C. and {Smith}, M. and {Sozzetti}, A. and {Utrilla}, E. and {van Leeuwen}, M. and {Abbas}, U. and {{\'A}brah{\'a}m}, P. and {Abreu Aramburu}, A. and {Aerts}, C. and {Aguado}, J.~J. and {Ajaj}, M. and {Aldea-Montero}, F. and {Altavilla}, G. and {{\'A}lvarez}, M.~A. and {Alves}, J. and {Anders}, F. and {Anderson}, R.~I. and {Anglada Varela}, E. and {Antoja}, T. and {Baines}, D. and {Baker}, S.~G. and {Balaguer-N{\'u}{\~n}ez}, L. and {Balbinot}, E. and {Balog}, Z. and {Barache}, C. and {Barbato}, D. and {Barros}, M. and {Barstow}, M.~A. and {Bartolom{\'e}}, S. and {Bassilana}, J. -L. and {Bauchet}, N. and {Becciani}, U. and {Bellazzini}, M. and {Berihuete}, A. and {Bernet}, M. and {Bertone}, S. and {Bianchi}, L. and {Binnenfeld}, A. and {Blanco-Cuaresma}, S. and {Blazere}, A. and {Boch}, T. and {Bombrun}, A. and {Bossini}, D. and {Bouquillon}, S. and {Bragaglia}, A. and {Bramante}, L. and {Breedt}, E. and {Bressan}, A. and {Brouillet}, N. and {Brugaletta}, E. and {Bucciarelli}, B. and {Burlacu}, A. and {Butkevich}, A.~G. and {Buzzi}, R. and {Caffau}, E. and {Cancelliere}, R. and {Cantat-Gaudin}, T. and {Carballo}, R. and {Carlucci}, T. and {Carnerero}, M.~I. and {Carrasco}, J.~M. and {Casamiquela}, L. and {Castellani}, M. and {Castro-Ginard}, A. and {Chaoul}, L. and {Charlot}, P. and {Chemin}, L. and {Chiaramida}, V. and {Chiavassa}, A. and {Chornay}, N. and {Comoretto}, G. and {Contursi}, G. and {Cooper}, W.~J. and {Cornez}, T. and {Cowell}, S. and {Crifo}, F. and {Cropper}, M. and {Crosta}, M. and {Crowley}, C. and {Dafonte}, C. and {Dapergolas}, A. and {David}, M. and {David}, P. and {de Laverny}, P. and {De Luise}, F. and {De March}, R. and {De Ridder}, J. and {de Souza}, R. and {de Torres}, A. and {del Peloso}, E.~F. and {del Pozo}, E. and {Delbo}, M. and {Delgado}, A. and {Delisle}, J. -B. and {Demouchy}, C. and {Dharmawardena}, T.~E. and {Di Matteo}, P. and {Diakite}, S. and {Diener}, C. and {Distefano}, E. and {Dolding}, C. and {Edvardsson}, B. and {Enke}, H. and {Fabre}, C. and {Fabrizio}, M. and {Faigler}, S. and {Fedorets}, G. and {Fernique}, P. and {Fienga}, A. and {Figueras}, F. and {Fournier}, Y. and {Fouron}, C. and {Fragkoudi}, F. and {Gai}, M. and {Garcia-Gutierrez}, A. and {Garcia-Reinaldos}, M. and {Garc{\'\i}a-Torres}, M. and {Garofalo}, A. and {Gavel}, A. and {Gavras}, P. and {Gerlach}, E. and {Geyer}, R. and {Giacobbe}, P. and {Gilmore}, G. and {Girona}, S. and {Giuffrida}, G. and {Gomel}, R. and {Gomez}, A. and {Gonz{\'a}lez-N{\'u}{\~n}ez}, J. and {Gonz{\'a}lez-Santamar{\'\i}a}, I. and {Gonz{\'a}lez-Vidal}, J.~J. and {Granvik}, M. and {Guillout}, P. and {Guiraud}, J. and {Guti{\'e}rrez-S{\'a}nchez}, R. and {Guy}, L.~P. and {Hatzidimitriou}, D. and {Hauser}, M. and {Haywood}, M. and {Helmer}, A. and {Helmi}, A. and {Sarmiento}, M.~H. and {Hidalgo}, S.~L. and {Hilger}, T. and {H{\l}adczuk}, N. and {Hobbs}, D. and {Holland}, G. and {Huckle}, H.~E. and {Jardine}, K. and {Jasniewicz}, G. and {Jean-Antoine Piccolo}, A. and {Jim{\'e}nez-Arranz}, {\'O}. and {Jorissen}, A. and {Juaristi Campillo}, J. and {Julbe}, F. and {Karbevska}, L. and {Kervella}, P. and {Khanna}, S. and {Kontizas}, M. and {Kordopatis}, G. and {Korn}, A.~J. and {K{\'o}sp{\'a}l}, {\'A}. and {Kostrzewa-Rutkowska}, Z. and {Kruszy{\'n}ska}, K. and {Kun}, M. and {Laizeau}, P. and {Lambert}, S. and {Lanza}, A.~F. and {Lasne}, Y. and {Le Campion}, J. -F. and {Lebreton}, Y. and {Lebzelter}, T. and {Leccia}, S. and {Leclerc}, N. and {Lecoeur-Taibi}, I. and {Liao}, S. and {Licata}, E.~L. and {Lindstr{\o}m}, H.~E.~P. and {Lister}, T.~A. and {Livanou}, E. and {Lobel}, A. and {Lorca}, A. and {Loup}, C. and {Madrero Pardo}, P. and {Magdaleno Romeo}, A. and {Managau}, S. and {Mann}, R.~G. and {Manteiga}, M. and {Marchant}, J.~M. and {Marconi}, M. and {Marcos}, J. and {Marcos Santos}, M.~M.~S. and {Mar{\'\i}n Pina}, D. and {Marinoni}, S. and {Marocco}, F. and {Marshall}, D.~J. and {Martin Polo}, L. and {Mart{\'\i}n-Fleitas}, J.~M. and {Marton}, G. and {Mary}, N. and {Masip}, A. and {Massari}, D. and {Mastrobuono-Battisti}, A. and {Mazeh}, T. and {McMillan}, P.~J. and {Messina}, S. and {Michalik}, D. and {Millar}, N.~R. and {Mints}, A. and {Molina}, D. and {Molinaro}, R. and {Moln{\'a}r}, L. and {Monari}, G. and {Mongui{\'o}}, M. and {Montegriffo}, P. and {Montero}, A. and {Mor}, R. and {Mora}, A. and {Morbidelli}, R. and {Morel}, T. and {Morris}, D. and {Muraveva}, T. and {Murphy}, C.~P. and {Musella}, I. and {Nagy}, Z. and {Noval}, L. and {Oca{\~n}a}, F. and {Ogden}, A. and {Ordenovic}, C. and {Osinde}, J.~O. and {Pagani}, C. and {Pagano}, I. and {Palaversa}, L. and {Palicio}, P.~A. and {Pallas-Quintela}, L. and {Panahi}, A. and {Payne-Wardenaar}, S. and {Pe{\~n}alosa Esteller}, X. and {Penttil{\"a}}, A. and {Pichon}, B. and {Piersimoni}, A.~M. and {Pineau}, F. -X. and {Plachy}, E. and {Plum}, G. and {Poggio}, E. and {Pr{\v{s}}a}, A. and {Pulone}, L. and {Racero}, E. and {Ragaini}, S. and {Rainer}, M. and {Raiteri}, C.~M. and {Rambaux}, N. and {Ramos}, P. and {Ramos-Lerate}, M. and {Re Fiorentin}, P. and {Regibo}, S. and {Richards}, P.~J. and {Rios Diaz}, C. and {Ripepi}, V. and {Riva}, A. and {Rix}, H. -W. and {Rixon}, G. and {Robichon}, N. and {Robin}, A.~C. and {Robin}, C. and {Roelens}, M. and {Rogues}, H.~R.~O. and {Rohrbasser}, L. and {Romero-G{\'o}mez}, M. and {Rowell}, N. and {Royer}, F. and {Ruz Mieres}, D. and {Rybicki}, K.~A. and {Sadowski}, G. and {S{\'a}ez N{\'u}{\~n}ez}, A. and {Sagrist{\`a} Sell{\'e}s}, A. and {Sahlmann}, J. and {Salguero}, E. and {Samaras}, N. and {Sanchez Gimenez}, V. and {Sanna}, N. and {Santove{\~n}a}, R. and {Sarasso}, M. and {Schultheis}, M. and {Sciacca}, E. and {Segol}, M. and {Segovia}, J.~C. and {S{\'e}gransan}, D. and {Semeux}, D. and {Shahaf}, S. and {Siddiqui}, H.~I. and {Siebert}, A. and {Siltala}, L. and {Silvelo}, A. and {Slezak}, E. and {Slezak}, I. and {Smart}, R.~L. and {Snaith}, O.~N. and {Solano}, E. and {Solitro}, F. and {Souami}, D. and {Souchay}, J. and {Spagna}, A. and {Spina}, L. and {Spoto}, F. and {Steele}, I.~A. and {Steidelm{\"u}ller}, H. and {Stephenson}, C.~A. and {S{\"u}veges}, M. and {Surdej}, J. and {Szabados}, L. and {Szegedi-Elek}, E. and {Taris}, F. and {Taylor}, M.~B. and {Teixeira}, R. and {Tolomei}, L. and {Tonello}, N. and {Torra}, F. and {Torra}, J. and {Torralba Elipe}, G. and {Trabucchi}, M. and {Tsounis}, A.~T. and {Turon}, C. and {Ulla}, A. and {Unger}, N. and {Vaillant}, M.~V. and {van Dillen}, E. and {van Reeven}, W. and {Vanel}, O. and {Vecchiato}, A. and {Viala}, Y. and {Vicente}, D. and {Voutsinas}, S. and {Weiler}, M. and {Wevers}, T. and {Wyrzykowski}, {\L}. and {Yoldas}, A. and {Yvard}, P. and {Zhao}, H. and {Zorec}, J. and {Zucker}, S. and {Zwitter}, T.},
        title = "{Gaia Data Release 3. Summary of the content and survey properties}",
      journal = {\aap},
         year = 2023,
        month = jun,
       volume = {674},
          eid = {A1},
        pages = {A1},
          doi = {10.1051/0004-6361/202243940},
archivePrefix = {arXiv},
       eprint = {2208.00211},
 primaryClass = {astro-ph.GA},
       adsurl = {https://ui.adsabs.harvard.edu/abs/2023A&A...674A...1G}
}

@ARTICLE{GaiaCol_2018_DR2_survey,
       author = {{Gaia Collaboration} and {Brown}, A.~G.~A. and {Vallenari}, A. and {Prusti}, T. and {de Bruijne}, J.~H.~J. and {Babusiaux}, C. and {Bailer-Jones}, C.~A.~L. and {Biermann}, M. and {Evans}, D.~W. and {Eyer}, L. and {Jansen}, F. and {Jordi}, C. and {Klioner}, S.~A. and {Lammers}, U. and {Lindegren}, L. and {Luri}, X. and {Mignard}, F. and {Panem}, C. and {Pourbaix}, D. and {Randich}, S. and {Sartoretti}, P. and {Siddiqui}, H.~I. and {Soubiran}, C. and {van Leeuwen}, F. and {Walton}, N.~A. and {Arenou}, F. and {Bastian}, U. and {Cropper}, M. and {Drimmel}, R. and {Katz}, D. and {Lattanzi}, M.~G. and {Bakker}, J. and {Cacciari}, C. and {Casta{\~n}eda}, J. and {Chaoul}, L. and {Cheek}, N. and {De Angeli}, F. and {Fabricius}, C. and {Guerra}, R. and {Holl}, B. and {Masana}, E. and {Messineo}, R. and {Mowlavi}, N. and {Nienartowicz}, K. and {Panuzzo}, P. and {Portell}, J. and {Riello}, M. and {Seabroke}, G.~M. and {Tanga}, P. and {Th{\'e}venin}, F. and {Gracia-Abril}, G. and {Comoretto}, G. and {Garcia-Reinaldos}, M. and {Teyssier}, D. and {Altmann}, M. and {Andrae}, R. and {Audard}, M. and {Bellas-Velidis}, I. and {Benson}, K. and {Berthier}, J. and {Blomme}, R. and {Burgess}, P. and {Busso}, G. and {Carry}, B. and {Cellino}, A. and {Clementini}, G. and {Clotet}, M. and {Creevey}, O. and {Davidson}, M. and {De Ridder}, J. and {Delchambre}, L. and {Dell'Oro}, A. and {Ducourant}, C. and {Fern{\'a}ndez-Hern{\'a}ndez}, J. and {Fouesneau}, M. and {Fr{\'e}mat}, Y. and {Galluccio}, L. and {Garc{\'\i}a-Torres}, M. and {Gonz{\'a}lez-N{\'u}{\~n}ez}, J. and {Gonz{\'a}lez-Vidal}, J.~J. and {Gosset}, E. and {Guy}, L.~P. and {Halbwachs}, J. -L. and {Hambly}, N.~C. and {Harrison}, D.~L. and {Hern{\'a}ndez}, J. and {Hestroffer}, D. and {Hodgkin}, S.~T. and {Hutton}, A. and {Jasniewicz}, G. and {Jean-Antoine-Piccolo}, A. and {Jordan}, S. and {Korn}, A.~J. and {Krone-Martins}, A. and {Lanzafame}, A.~C. and {Lebzelter}, T. and {L{\"o}ffler}, W. and {Manteiga}, M. and {Marrese}, P.~M. and {Mart{\'\i}n-Fleitas}, J.~M. and {Moitinho}, A. and {Mora}, A. and {Muinonen}, K. and {Osinde}, J. and {Pancino}, E. and {Pauwels}, T. and {Petit}, J. -M. and {Recio-Blanco}, A. and {Richards}, P.~J. and {Rimoldini}, L. and {Robin}, A.~C. and {Sarro}, L.~M. and {Siopis}, C. and {Smith}, M. and {Sozzetti}, A. and {S{\"u}veges}, M. and {Torra}, J. and {van Reeven}, W. and {Abbas}, U. and {Abreu Aramburu}, A. and {Accart}, S. and {Aerts}, C. and {Altavilla}, G. and {{\'A}lvarez}, M.~A. and {Alvarez}, R. and {Alves}, J. and {Anderson}, R.~I. and {Andrei}, A.~H. and {Anglada Varela}, E. and {Antiche}, E. and {Antoja}, T. and {Arcay}, B. and {Astraatmadja}, T.~L. and {Bach}, N. and {Baker}, S.~G. and {Balaguer-N{\'u}{\~n}ez}, L. and {Balm}, P. and {Barache}, C. and {Barata}, C. and {Barbato}, D. and {Barblan}, F. and {Barklem}, P.~S. and {Barrado}, D. and {Barros}, M. and {Barstow}, M.~A. and {Bartholom{\'e} Mu{\~n}oz}, S. and {Bassilana}, J. -L. and {Becciani}, U. and {Bellazzini}, M. and {Berihuete}, A. and {Bertone}, S. and {Bianchi}, L. and {Bienaym{\'e}}, O. and {Blanco-Cuaresma}, S. and {Boch}, T. and {Boeche}, C. and {Bombrun}, A. and {Borrachero}, R. and {Bossini}, D. and {Bouquillon}, S. and {Bourda}, G. and {Bragaglia}, A. and {Bramante}, L. and {Breddels}, M.~A. and {Bressan}, A. and {Brouillet}, N. and {Br{\"u}semeister}, T. and {Brugaletta}, E. and {Bucciarelli}, B. and {Burlacu}, A. and {Busonero}, D. and {Butkevich}, A.~G. and {Buzzi}, R. and {Caffau}, E. and {Cancelliere}, R. and {Cannizzaro}, G. and {Cantat-Gaudin}, T. and {Carballo}, R. and {Carlucci}, T. and {Carrasco}, J.~M. and {Casamiquela}, L. and {Castellani}, M. and {Castro-Ginard}, A. and {Charlot}, P. and {Chemin}, L. and {Chiavassa}, A. and {Cocozza}, G. and {Costigan}, G. and {Cowell}, S. and {Crifo}, F. and {Crosta}, M. and {Crowley}, C. and {Cuypers}, J. and {Dafonte}, C. and {Damerdji}, Y. and {Dapergolas}, A. and {David}, P. and {David}, M. and {de Laverny}, P. and {De Luise}, F. and {De March}, R. and {de Martino}, D. and {de Souza}, R. and {de Torres}, A. and {Debosscher}, J. and {del Pozo}, E. and {Delbo}, M. and {Delgado}, A. and {Delgado}, H.~E. and {Di Matteo}, P. and {Diakite}, S. and {Diener}, C. and {Distefano}, E. and {Dolding}, C. and {Drazinos}, P. and {Dur{\'a}n}, J. and {Edvardsson}, B. and {Enke}, H. and {Eriksson}, K. and {Esquej}, P. and {Eynard Bontemps}, G. and {Fabre}, C. and {Fabrizio}, M. and {Faigler}, S. and {Falc{\~a}o}, A.~J. and {Farr{\`a}s Casas}, M. and {Federici}, L. and {Fedorets}, G. and {Fernique}, P. and {Figueras}, F. and {Filippi}, F. and {Findeisen}, K. and {Fonti}, A. and {Fraile}, E. and {Fraser}, M. and {Fr{\'e}zouls}, B. and {Gai}, M. and {Galleti}, S. and {Garabato}, D. and {Garc{\'\i}a-Sedano}, F. and {Garofalo}, A. and {Garralda}, N. and {Gavel}, A. and {Gavras}, P. and {Gerssen}, J. and {Geyer}, R. and {Giacobbe}, P. and {Gilmore}, G. and {Girona}, S. and {Giuffrida}, G. and {Glass}, F. and {Gomes}, M. and {Granvik}, M. and {Gueguen}, A. and {Guerrier}, A. and {Guiraud}, J. and {Guti{\'e}rrez-S{\'a}nchez}, R. and {Haigron}, R. and {Hatzidimitriou}, D. and {Hauser}, M. and {Haywood}, M. and {Heiter}, U. and {Helmi}, A. and {Heu}, J. and {Hilger}, T. and {Hobbs}, D. and {Hofmann}, W. and {Holland}, G. and {Huckle}, H.~E. and {Hypki}, A. and {Icardi}, V. and {Jan{\ss}en}, K. and {Jevardat de Fombelle}, G. and {Jonker}, P.~G. and {Juh{\'a}sz}, {\'A}. L. and {Julbe}, F. and {Karampelas}, A. and {Kewley}, A. and {Klar}, J. and {Kochoska}, A. and {Kohley}, R. and {Kolenberg}, K. and {Kontizas}, M. and {Kontizas}, E. and {Koposov}, S.~E. and {Kordopatis}, G. and {Kostrzewa-Rutkowska}, Z. and {Koubsky}, P. and {Lambert}, S. and {Lanza}, A.~F. and {Lasne}, Y. and {Lavigne}, J. -B. and {Le Fustec}, Y. and {Le Poncin-Lafitte}, C. and {Lebreton}, Y. and {Leccia}, S. and {Leclerc}, N. and {Lecoeur-Taibi}, I. and {Lenhardt}, H. and {Leroux}, F. and {Liao}, S. and {Licata}, E. and {Lindstr{\o}m}, H.~E.~P. and {Lister}, T.~A. and {Livanou}, E. and {Lobel}, A. and {L{\'o}pez}, M. and {Managau}, S. and {Mann}, R.~G. and {Mantelet}, G. and {Marchal}, O. and {Marchant}, J.~M. and {Marconi}, M. and {Marinoni}, S. and {Marschalk{\'o}}, G. and {Marshall}, D.~J. and {Martino}, M. and {Marton}, G. and {Mary}, N. and {Massari}, D. and {Matijevi{\v{c}}}, G. and {Mazeh}, T. and {McMillan}, P.~J. and {Messina}, S. and {Michalik}, D. and {Millar}, N.~R. and {Molina}, D. and {Molinaro}, R. and {Moln{\'a}r}, L. and {Montegriffo}, P. and {Mor}, R. and {Morbidelli}, R. and {Morel}, T. and {Morris}, D. and {Mulone}, A.~F. and {Muraveva}, T. and {Musella}, I. and {Nelemans}, G. and {Nicastro}, L. and {Noval}, L. and {O'Mullane}, W. and {Ord{\'e}novic}, C. and {Ord{\'o}{\~n}ez-Blanco}, D. and {Osborne}, P. and {Pagani}, C. and {Pagano}, I. and {Pailler}, F. and {Palacin}, H. and {Palaversa}, L. and {Panahi}, A. and {Pawlak}, M. and {Piersimoni}, A.~M. and {Pineau}, F. -X. and {Plachy}, E. and {Plum}, G. and {Poggio}, E. and {Poujoulet}, E. and {Pr{\v{s}}a}, A. and {Pulone}, L. and {Racero}, E. and {Ragaini}, S. and {Rambaux}, N. and {Ramos-Lerate}, M. and {Regibo}, S. and {Reyl{\'e}}, C. and {Riclet}, F. and {Ripepi}, V. and {Riva}, A. and {Rivard}, A. and {Rixon}, G. and {Roegiers}, T. and {Roelens}, M. and {Romero-G{\'o}mez}, M. and {Rowell}, N. and {Royer}, F. and {Ruiz-Dern}, L. and {Sadowski}, G. and {Sagrist{\`a} Sell{\'e}s}, T. and {Sahlmann}, J. and {Salgado}, J. and {Salguero}, E. and {Sanna}, N. and {Santana-Ros}, T. and {Sarasso}, M. and {Savietto}, H. and {Schultheis}, M. and {Sciacca}, E. and {Segol}, M. and {Segovia}, J.~C. and {S{\'e}gransan}, D. and {Shih}, I. -C. and {Siltala}, L. and {Silva}, A.~F. and {Smart}, R.~L. and {Smith}, K.~W. and {Solano}, E. and {Solitro}, F. and {Sordo}, R. and {Soria Nieto}, S. and {Souchay}, J. and {Spagna}, A. and {Spoto}, F. and {Stampa}, U. and {Steele}, I.~A. and {Steidelm{\"u}ller}, H. and {Stephenson}, C.~A. and {Stoev}, H. and {Suess}, F.~F. and {Surdej}, J. and {Szabados}, L. and {Szegedi-Elek}, E. and {Tapiador}, D. and {Taris}, F. and {Tauran}, G. and {Taylor}, M.~B. and {Teixeira}, R. and {Terrett}, D. and {Teyssandier}, P. and {Thuillot}, W. and {Titarenko}, A. and {Torra Clotet}, F. and {Turon}, C. and {Ulla}, A. and {Utrilla}, E. and {Uzzi}, S. and {Vaillant}, M. and {Valentini}, G. and {Valette}, V. and {van Elteren}, A. and {Van Hemelryck}, E. and {van Leeuwen}, M. and {Vaschetto}, M. and {Vecchiato}, A. and {Veljanoski}, J. and {Viala}, Y. and {Vicente}, D. and {Vogt}, S. and {von Essen}, C. and {Voss}, H. and {Votruba}, V. and {Voutsinas}, S. and {Walmsley}, G. and {Weiler}, M. and {Wertz}, O. and {Wevers}, T. and {Wyrzykowski}, {\L}. and {Yoldas}, A. and {{\v{Z}}erjal}, M. and {Ziaeepour}, H. and {Zorec}, J. and {Zschocke}, S. and {Zucker}, S. and {Zurbach}, C. and {Zwitter}, T.},
        title = "{Gaia Data Release 2. Summary of the contents and survey properties}",
      journal = {\aap},
         year = 2018,
        month = aug,
       volume = {616},
          eid = {A1},
        pages = {A1},
          doi = {10.1051/0004-6361/201833051},
archivePrefix = {arXiv},
       eprint = {1804.09365},
 primaryClass = {astro-ph.GA},
       adsurl = {https://ui.adsabs.harvard.edu/abs/2018A&A...616A...1G}
}

@ARTICLE{Iorio2021,
       author = {{Iorio}, Giuliano and {Belokurov}, Vasily},
        title = "{Chemo-kinematics of the Gaia RR Lyrae: the halo and the disc}",
      journal = {MNRAS},
         year = 2021,
        month = apr,
       volume = {502},
       number = {4},
        pages = {5686-5710},
          doi = {10.1093/mnras/stab005},
archivePrefix = {arXiv},
       eprint = {2008.02280},
 primaryClass = {astro-ph.GA},
       adsurl = {https://ui.adsabs.harvard.edu/abs/2021MNRAS.502.5686I}
}

@ARTICLE{Clementini2023_SOS_GaiaDR3,
       author = {{Clementini}, G. and {Ripepi}, V. and {Garofalo}, A. and {Molinaro}, R. and {Muraveva}, T. and {Leccia}, S. and {Rimoldini}, L. and {Holl}, B. and {Jevardat de Fombelle}, G. and {Sartoretti}, P. and {Marchal}, O. and {Audard}, M. and {Nienartowicz}, K. and {Andrae}, R. and {Marconi}, M. and {Szabados}, L. and {Evans}, D.~W. and {Lecoeur-Taibi}, I. and {Mowlavi}, N. and {Musella}, I. and {Eyer}, L.},
        title = "{Gaia Data Release 3. Specific processing and validation of all-sky RR Lyrae and Cepheid stars: The RR Lyrae sample}",
      journal = {\aap},
         year = 2023,
        month = jun,
       volume = {674},
          eid = {A18},
        pages = {A18},
          doi = {10.1051/0004-6361/202243964},
archivePrefix = {arXiv},
       eprint = {2206.06278},
 primaryClass = {astro-ph.SR},
       adsurl = {https://ui.adsabs.harvard.edu/abs/2023A&A...674A..18C}
}

@ARTICLE{Muraveva2018,
       author = {{Muraveva}, Tatiana and {Delgado}, Hector E. and {Clementini}, Gisella and {Sarro}, Luis M. and {Garofalo}, Alessia},
        title = "{RR Lyrae stars as standard candles in the Gaia Data Release 2 Era}",
      journal = {\mnras},
         year = 2018,
        month = nov,
       volume = {481},
       number = {1},
        pages = {1195-1211},
          doi = {10.1093/mnras/sty2241},
archivePrefix = {arXiv},
       eprint = {1805.08742},
 primaryClass = {astro-ph.SR},
       adsurl = {https://ui.adsabs.harvard.edu/abs/2018MNRAS.481.1195M}
}

@ARTICLE{Garofalo2022,
       author = {{Garofalo}, A. and {Delgado}, H.~E. and {Sarro}, L.~M. and {Clementini}, G. and {Muraveva}, T. and {Marconi}, M. and {Ripepi}, V.},
        title = "{New LZ and PW(Z) relations of RR Lyrae stars calibrated with Gaia EDR3 parallaxes}",
      journal = {\mnras},
         year = 2022,
        month = jun,
       volume = {513},
       number = {1},
        pages = {788-806},
          doi = {10.1093/mnras/stac735},
archivePrefix = {arXiv},
       eprint = {2203.07435},
 primaryClass = {astro-ph.SR},
       adsurl = {https://ui.adsabs.harvard.edu/abs/2022MNRAS.513..788G}
}

@article{Sarbadhicary_2021,
doi = {10.3847/1538-4357/abca86},
url = {https://dx.doi.org/10.3847/1538-4357/abca86},
year = {2021},
month = {may},
publisher = {The American Astronomical Society},
volume = {912},
number = {2},
pages = {140},
author = {Sumit K. Sarbadhicary and Mairead Heiger and Carles Badenes and Cecilia Mateu and Jeffrey A. Newman and Robin Ciardullo and Na’ama Hallakoun and Dan Maoz and Laura Chomiuk},
title = {The RR Lyrae Delay-time Distribution: A Novel Perspective on Models of Old Stellar Populations},
journal = {The Astrophysical Journal},
}

@ARTICLE{Pietrzynski2012,
       author = {{Pietrzy{\'n}ski}, G. and {Thompson}, I.~B. and {Gieren}, W. and {Graczyk}, D. and {St{\k{e}}pie{\'n}}, K. and {Bono}, G. and {Moroni}, P.~G. Prada and {Pilecki}, B. and {Udalski}, A. and {Soszy{\'n}ski}, I. and {Preston}, G.~W. and {Nardetto}, N. and {McWilliam}, A. and {Roederer}, I.~U. and {G{\'o}rski}, M. and {Konorski}, P. and {Storm}, J.},
        title = "{RR-Lyrae-type pulsations from a 0.26-solar-mass star in a binary system}",
      journal = {\nat},
         year = 2012,
        month = apr,
       volume = {484},
       number = {7392},
        pages = {75-77},
          doi = {10.1038/nature10966},
archivePrefix = {arXiv},
       eprint = {1204.1872},
 primaryClass = {astro-ph.SR},
       adsurl = {https://ui.adsabs.harvard.edu/abs/2012Natur.484...75P}
}

@ARTICLE{Gratton2010,
       author = {{Gratton}, R.~G. and {Carretta}, E. and {Bragaglia}, A. and {Lucatello}, S. and {D'Orazi}, V.},
        title = "{The second and third parameters of the horizontal branch in globular clusters}",
      journal = {\aap},
         year = 2010,
        month = jul,
       volume = {517},
          eid = {A81},
        pages = {A81},
          doi = {10.1051/0004-6361/200912572},
archivePrefix = {arXiv},
       eprint = {1004.3862},
 primaryClass = {astro-ph.SR},
       adsurl = {https://ui.adsabs.harvard.edu/abs/2010A&A...517A..81G}
}

\begin{appendix}
\onecolumn

\section{Discarded stars}\label{a:discarded}

\begin{figure*}[ht!] 
    \includegraphics[width=0.9\textwidth]{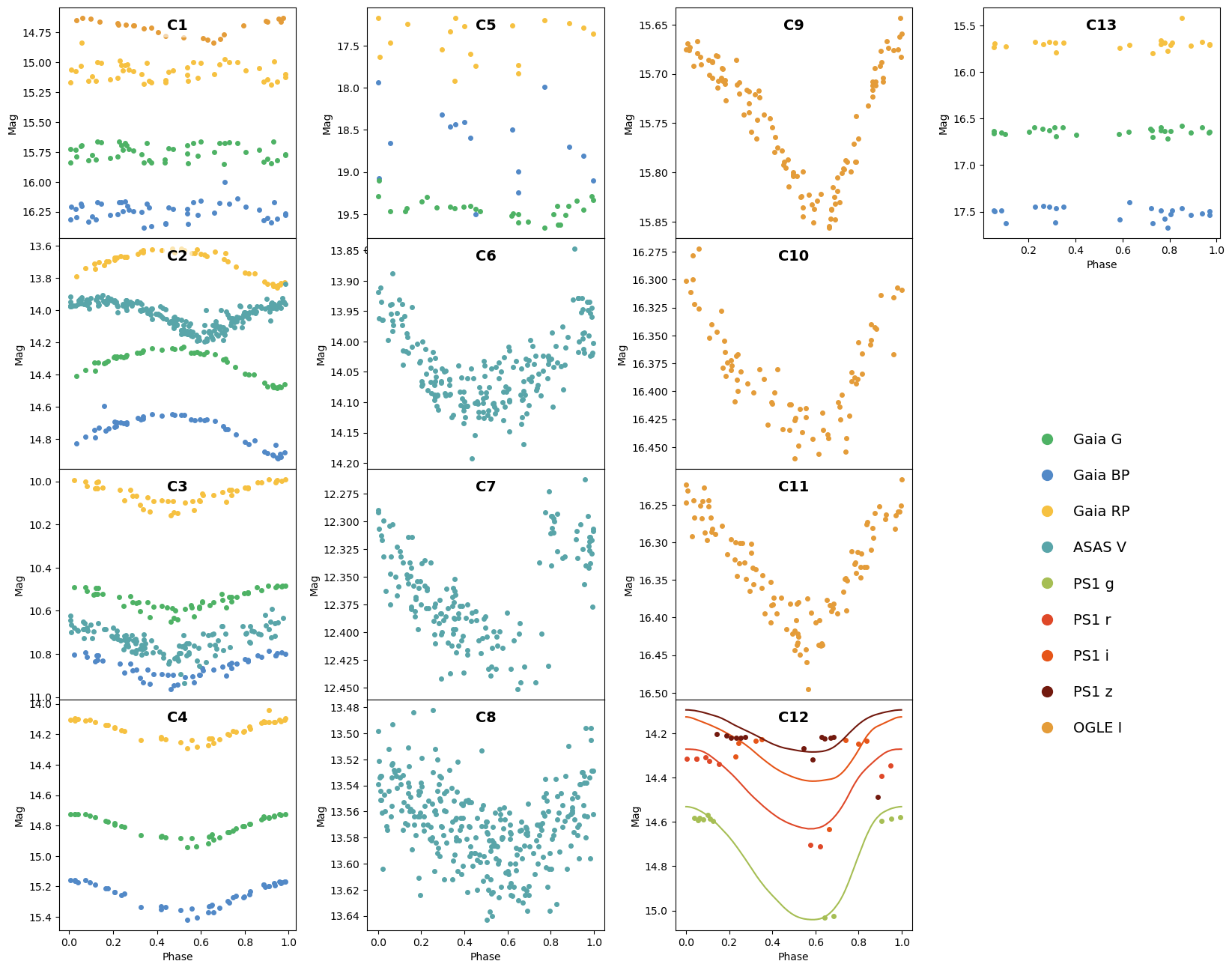}
    \caption{Phase folded light curves for each of the 13 out of 14 stars discarded from the sample, with available time series data. Star  3330383474578837248 is not shown as it was identified by ZTF alone and its time series data is not publicly available, not does it have time series information in \Gaia\ DR3.}
    \label{fa:discarded_lcs}
\end{figure*}

\begin{figure*}[ht!] 
    \centering
    \includegraphics[width=0.75\textwidth]{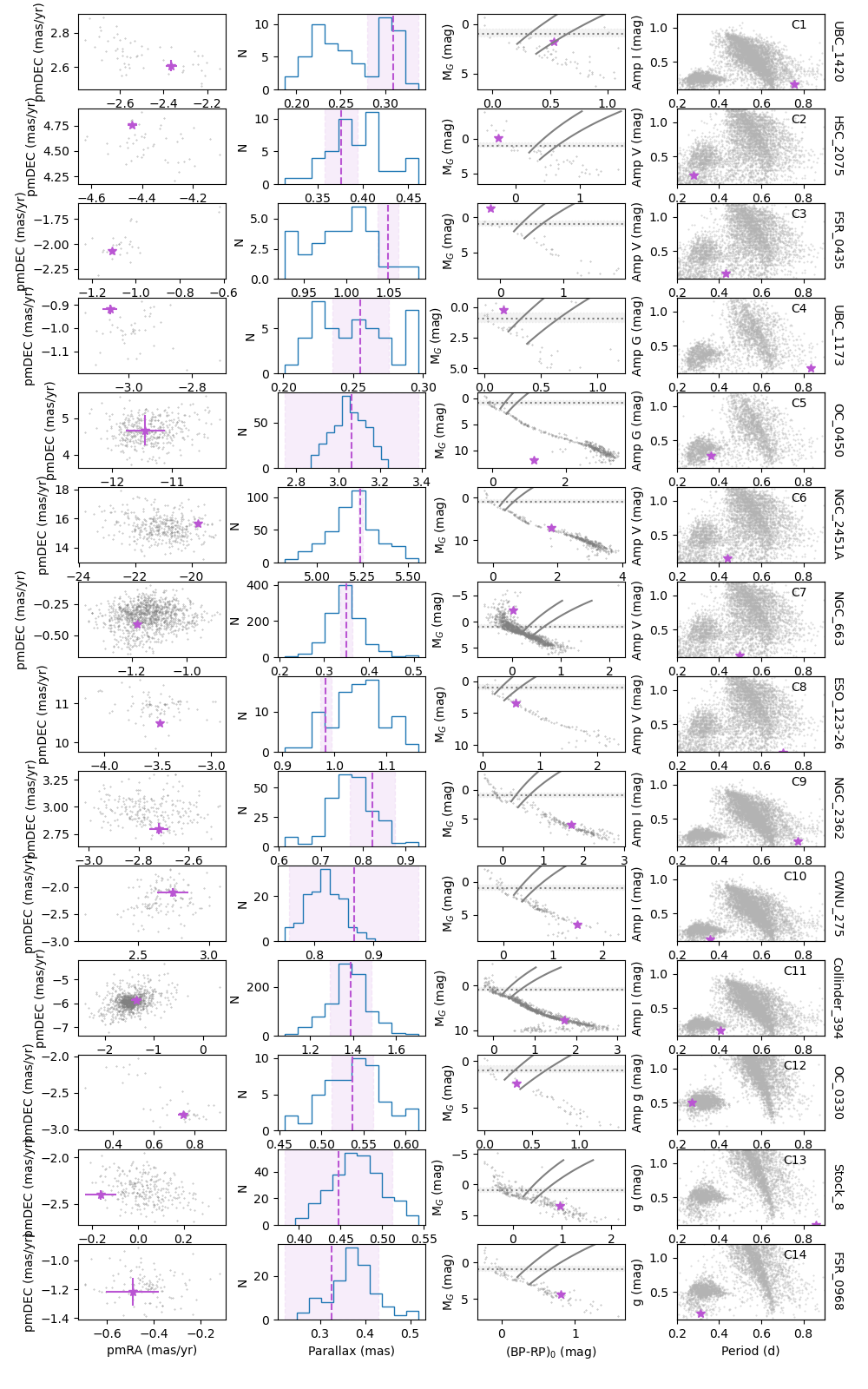}
    \caption{Discarded RRL stars (purple star) and cluster members (light gray) are shown, from left to right, in the plane of proper motions (pmDEC versus pmRA), parallax, \Gaia\ CAMD and Period-Amplitude diagram. In the CAMD, the solid lines correspond to IS limits from \citet{Marconi2015}; the dotted line shows the expected location of the HB at each cluster's metallicity from \citet{Garofalo2022}, and the absolute magnitude and intrinsic colours computed based on the distance and extinction according to \citet{Hunt2024}; the shading represents the uncertainty in the HB luminosity corresponding to $\Delta\FeH=0.5$~dex. The star's ID from Table~\ref{ta:discarded_lcinfo} and the cluster's name are shown in the fourth panel top right corner and Y axis, respectively.}
    \label{fa:discarded_4d}
\end{figure*}

The search for RRL members in OCs yielded an initial sample of 15 stars. Here we discuss the 14 stars which we discarded as misclassified RRLs, mainly based on the appearance of their phase-folded light curves and/or position in the CAMD.  Figure~\ref{fa:discarded_lcs} shows the phase-folded light curves for all the time series publicly available from \Gaia, ASAS-SN, PS1 and OGLE-IV for these stars. In this plot, stars C13 and C14 were identified as RRLs by ZTF alone. Since the ZTF time series are not available,  we show the \Gaia~DR3 for C13. We were unable to find any time series information for star C14, hence it is not shown in Fig.~\ref{fa:discarded_lcs}. Figure~\ref{fa:discarded_4d} shows the proper motion plane, parallax histogram, CAMD and Period-Amplitude diagram for the star of interest (purple star) and cluster members from \citet{Hunt2023}, similarly to Fig.~\ref{f:5d}.
Table~\ref{ta:discarded_lcinfo} summarises the available light curve information for the discarded stars.

Since the RRLs we are interested in could belong to unresolved binaries we should not, a priori, rule out stars that lie outside the IS. We can, however, discard any stars below the expected HB luminosity since an unresolved binary can only be more luminous than the RRL, assuming it has a `normal' luminosity. The position in the CAMD of Fig.~\ref{fa:discarded_4d} for several stars (e.g. C5, C6, etc.) places them well below the HB and, in many cases, directly on the MS. These cases are denoted as below-HB/MS in the comments of Table~\ref{ta:discarded_lcinfo} as reasons for the stars to be discarded from our sample. In what follows we describe the criteria used to discard each star, also summarised in Table~\ref{ta:discarded_lcinfo}. 

Stars C1--C4, show the characteristic light curves of contact or semi-detached eclipsing binaries (W~UMa/EW and $\beta$Lyr/E$\beta$), consistent with their position on or near the MS as seen in the CAMD of Figure~\ref{fa:discarded_4d}. Star C5 has a very noisy light curve and is placed well below the MS. Stars C6, C8, C9, C10 and C11 have reasonably well behaved light curves, but too short amplitudes much more consistent with being either $\delta$~Scuti stars or eclipsing binary (EB), the latter being more consistent with their position in the MS but outside the IS, in the CAMD. Star C7 is bluer than the IS and more luminous than the HB, which would make it a good candidate, except for its extremely low amplitude, particularly for its period, and the significant gap in its light curve. This star might be worth revisiting, but for now it is discarded as a possible EB.
Stars C12 and C13 show almost constant light curves with a few deviated points, usually characteristic of detached eclipsing binaries (Algol/EA). This is indicated with the corresponding variability classification in the comments of Table~\ref{ta:discarded_lcinfo}. Finally, star C14 has a very short amplitude and is located right at the MS, well below the HB. Although it is the only star without time series data available, its very short amplitudes and location in the CAMD make it likely to be an EB.

\begin{table}[ht!] 
    \caption{Stars discarded as misclassified RRLs.}

    \centering
    \setlength{\tabcolsep}{4pt} 
    \begin{tabular}{ccccc}
\hline
\hline
ID & Gaia Source ID & Periods (d) & Amplitudes (mag) & Comments \\
 & & (G,A,O) & (BP,RP,G,V,I) & \\
\hline
\hline
C1 & 5594538779611595136 & (--,--, 0.7541) & (--,--,--,--,   0.18) & EW, MS\\
C2 & 5533487296963447552 & (--, 0.2792,--) & (--,--,--,   0.23,--)& EW, MS\\
C3 & 2012045019718073728 & (--, 0.4310,--) & (--,--,--,   0.18,--) & EW, MS\\
C4 & 2004077065120273664 & ( 0.8335,--,--) & (   0.21,   0.16,   0.18,--,--)& EW, MS\\
C5 & 5534158205211222144 & ( 0.3622,--,--) & (   0.80,   0.27,   0.27,--,--)& noisy, below-HB\\
C6 & 5585551027962400384 & (--, 0.4400,--) & (--,--,--,   0.17,--) & EB, MS\\
C7 & 511219584907760896 & (--, 0.4992,--) & (--,--,--,   0.13,--) & off-MS, short Amp\\
C8 & 5294043694838471552 & (--, 0.7017,--) & (--,--,--,   0.09,--) & EB, MS\\
C9 & 5616530661434711680 & (--,--, 0.7728) & (--,--,--,--,   0.17) & EB, MS\\
C10 & 5957810991851847296 & (--,--, 0.3583) & (--,--,--,--,   0.13) & EB, MS\\
C11 & 4085159414905583104 & (--,--, 0.4053) & (--,--,--,--,   0.17) & EB, MS\\
\hline
 & & (PS1) & ($g$,$r$,$i$,$z$)\\
\hline 
C12 & 3355237350809737344 & (0.2723) & (0.51,   0.36,   0.29,   0.19) & constant/EA?\\
\hline
 & & (ZTF) & ($g$,$r$)\\ 
\hline 
C13 & 182831054077294336 & (0.8605) & (0.10,0.10) & constant/EA?\\
C14 & 3330383474578837248 & (0.3134) & (0.19,0.12) & MS, below-HB\\
\hline 
\hline
    \end{tabular}
    \label{ta:discarded_lcinfo}
\end{table}

\section{Extinction}\label{a:Av}

\begin{figure}[ht!] 
    \centering
    \includegraphics[width=0.8\columnwidth]{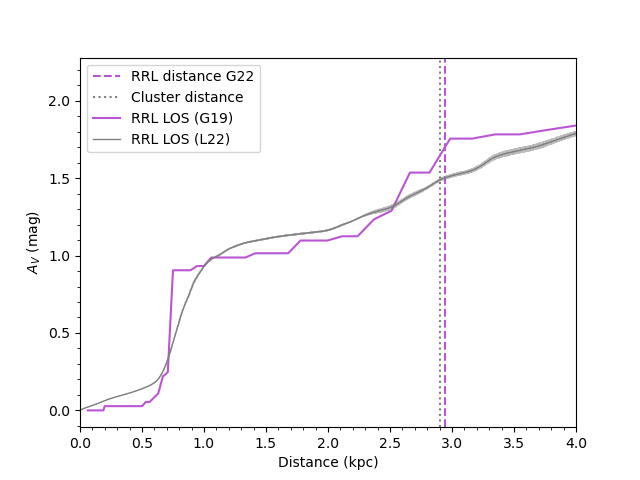}
    \caption{Extinction in the V band along the RRL's line of sight from the G19 and L22 extinction maps. The width of the L22 curve represents the $1\sigma$ uncertainty in extinction at fixed distance. The RRL's parallax distance and cluster's mean distance summarised in Table~\ref{t:astrom} are shown with the dashed and dotted lines respectively.}
    \label{fa:Av_vs_d}
\end{figure}

Trumpler~5 is located in a region with high differential extinction. In Figure~\ref{fa:Av_vs_d} we show the dependence of $A_V$ as a function of distance along the line of sight to the RRL, for the \citet{Lallement2022} and \citet{Green2019} 3D extinction maps. The overall trend is very similar but there are punctual differences up to $\sim0.25$~mag between the two maps. At the RRL's distance the G19 map predicts  larger extinction than the L22 map by $\sim0.15$~mag.

\end{appendix}

\end{document}